\documentclass[10pt, a4paper, twocolumn]{article}

\usepackage{amsmath, amssymb}
\usepackage{graphicx}
\usepackage{authblk}  
\usepackage{hyperref}

\usepackage{graphicx}
\usepackage{amsfonts}
\usepackage{amsthm}
\usepackage{cases}
\usepackage{wrapfig}
\usepackage{booktabs}
\usepackage{textcomp}
\usepackage{xcolor}
\usepackage{enumerate}
\usepackage{float}
\usepackage{caption}
\usepackage{lineno}
\usepackage{colortbl}

\usepackage{geometry}
\usepackage{algorithm}
\usepackage{algpseudocode}

\algnewcommand\algorithmicparallelfor{\textbf{parallel for}}
\algnewcommand\algorithmicendparallelfor{\textbf{end parallel for}}
\algblockdefx[ParallelFor]{ParallelFor}{EndParallelFor}%
[1]{\algorithmicparallelfor\ #1}%
{\algorithmicendparallelfor}

\usepackage[mathscr]{euscript}

\newtheorem{Remark}{Remark}

\usepackage{titlesec}

\titleformat{\section}[hang] 
{\normalfont\normalsize\bfseries} 
{\thesection}{0.4em}{} 

\titleformat{\subsection}[hang] 
{\normalfont\normalsize\bfseries} 
{\thesubsection}{0.4em}{} 

\usepackage{fancyhdr}  

\title{\textbf{{\Large{Deep Learning for Analyzing Chaotic Dynamics in Biological Time Series: Insights from Frog Heart Signals}}}}

\author[1,*]{{\large{Carmen Mayora-Cebollero}}}
\author[2]{{\large{Flavio H. Fenton}}}
\author[2,3]{{\large{Molly Halprin}}}
\author[2]{{\large{Conner Herndon}}}
\author[2]{{\large{\\Mikael J. Toye}}}
\author[1]{{\large{Roberto Barrio}}}

\affil[1]{{\small{IUMA, Computational Dynamics group and Department of Applied Mathematics, Universidad de Zaragoza, Zaragoza, Spain}}}
\affil[2]{{\small{School of Physics, Georgia Institute of Technology, Atlanta, USA}}}
\affil[3]{{\small{United States Patent and Trademark Office, Los Angeles, USA}}}

\affil[*]{{\small{Corresponding author. Carmen Mayora-Cebollero. cmayora@unizar.es}}}

\date{}  

\begin{document}
	

\twocolumn[{
		\maketitle
	\begin{center}
		\large \textbf{Abstract}
	\end{center}
	\noindent
		The study of experimental data is a relevant task in several physical, chemical and biological applications. In particular, the analysis of chaotic dynamics in cardiac systems is crucial as it can be related to some pathological arrhythmias. When working with short and noisy experimental time series, some standard techniques for chaos detection cannot provide reliable results because of such data characteristics. Moreover, when small datasets are available, Deep Learning techniques cannot be applied directly (that is, using part of the data to train the network, and using the trained network to analyze the remaining dataset). To avoid all these limitations, we propose an automatic algorithm that combines Deep Learning and some selection strategies based on a mathematical model of the same nature of the experimental data. To show its performance, we test it with experimental data obtained from ex-vivo frog heart experiments, obtaining highly accurate results.
			\vspace{0.5em} 
			
			\vspace{1em}
}]
	
\section{Introduction}

Nowadays, the analysis and study of experimental data is becoming more and more crucial to develop mathematical models and classify different possible behaviors. In data from real experiments, different features are common, such as different lengths in the time series, possibly a very short amount of data in most cases, or the impossibility of measuring all the variables of the system (such as latent variables). In this study, we focus on a first attempt to automate the analysis of chaos in cardiac time series (in our case, frog heart signals) using Deep Learning techniques and comparison with those of a periodically paced cardiac action potential map~\cite{hastings2000alternans,xie2007dispersion} to verify validity, as iterative maps remain a valuable tool for studying chaos in cardiac cell models~\cite{munoz2021controllability, stenzinger2023cardiac,wang2024intracellular}.  Chaos dynamics in the heart~\cite{Garfinkel1992,skinner1990chaos,710541} has been proposed to exist at two levels, in the ECG for healthy individuals~\cite{goldberger1991normal,poon1997decrease} and in tissue level during fibrillation~\cite{goldberger1986some,weiss1999chaos,gupta2021chaos} or during fast pacing~\cite{chialvo1987non,chialvo1990low}, as well as higher order periods~\cite{gizzi2013effects,iravanian2023complex}. Therefore, the methods presented here could help in the analysis and classification of chaos at different regimes as well as different animal species, including humans. Moreover, characterizing chaos in these contexts is essential for developing targeted strategies and designing more effective therapeutic interventions, particularly those based on chaos control techniques for the termination of arrhythmias \cite{lilienkamp2022taming,detal2022terminating,tyler2024experimental, ji2017synchronization}.

Various measures of complexity such as Lyapunov exponents~\cite{argyris1994exploration, wolf1985determining, rosenstein1993practical}, Permutation Entropy~\cite{BandtPompe02,Amigo10,stosic2022generalized,Riedl13}, and the 0\,-1 test for chaos~\cite{gottwald20160} are often used to perform chaos analysis (distinguish regular (e.g., periodic or quasiperiodic) and chaotic behavior) of dynamical systems~\cite{barrio2007three, EYEBEFOUDA2016259, barrio2021dynamical, halfar2020dynamical,shintani2024observation,islam2025dynamic,ma2025using,Toker20}. However, as these techniques are usually based on large time series, they sometimes present problems when used in real data as recordings are generally short and noisy~\cite{palus1998chaotic,zanin2021ordinal}. In some cases, preprocessing tasks can be useful to deal with such kind of datasets~\cite{kumar09,kumar25}. Moreover, when dealing with large experimental datasets it is necessary to have an automatic algorithm for chaos analysis since human intervention on the entire dataset is not feasible. Recently, some authors have used Deep Learning (Artificial Neural Networks) to detect chaos in a dynamical system~\cite{CGRV22, ChaosDetCody23, pathak2017using, mayora2024full}. In particular, in~\cite{mayora2024full}, it was illustrated that using Deep Learning just partial information, as single-variable time series (without any previous preprocessing technique), is enough to obtain a complete understanding of the dynamical behavior (approximation of the full Lyapunov exponents spectrum). Could Deep Learning (DL) be applied to analyze chaotic dynamics in an experimental dataset?  Notice that from the dynamical systems point of view a chaos analysis  consists of detecting if a time series has regular or chaotic behavior. From the DL perspective, it is a classification task. When applied to cardiac tissue data, such an approach could potentially enable the prediction and characterization of complex experimental dynamics~\cite{shahi2022machine,shahi2022prediction}.

\begin{figure}[h!]
	\begin{center}
		\includegraphics[width = 0.43\textwidth]{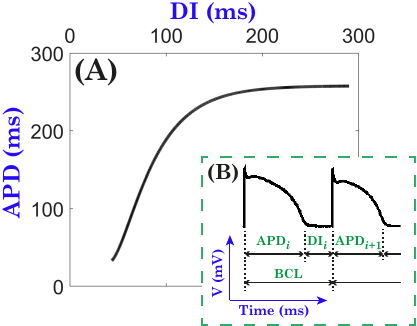}
		\caption{(A) APD restitution curve fitted to the kinetics of the Beeler-Reuter model. (B) Scheme to illustrate the APD, DI and BCL concepts.}
		\label{Scheme}
	\end{center}
\end{figure}

During pacing of experimental or simulated cardiac cell or tissue, trans-membrane voltage signals, called action potentials, can be obtained that look as shown in panel (B) of Figure~\ref{Scheme}. These signals can be characterized by an Action Potential Duration (APD), Diastolic Interval (DI), and Basic Cycle Length (BCL). The APD is the duration from the onset of cell depolarization to the completion of repolarization, given a certain threshold, where most commonly a value in between 75$\%$ to 90$\%$ of repolarization is used. The DI is then the time interval from the end of an APD to the onset of the next one, for the same threshold, which is basically the recovery time of the cell. The BCL is the time interval between consecutive sinus beats or the pacing cycle length used to stimulate the heart when experiments are carried out.

Using the APD and DI data, an APD restitution curve can be constructed to predict the APD given a particular period of stimulation. This is a one-dimensional map that has been shown useful to predict complex dynamics~\cite{nolasco1968graphic} and which has been shown experimentally to predict period-doubling bifurcations~\cite{guevara1984electrical}. Usually, it is a monotonically increasing function, but some experiments have shown that it could be biphasic~\cite{franz1988cycle,szigligeti1998intracellular}, which could lead to chaotic dynamics~\cite{watanabe1995biphasic,qu1997spatiotemporal,fenton2002multiple}. Mathematically, it is represented with a discrete equation of the form $\text{APD}_{i+1} = f(\text{BCL}-\text{APD}_{i}) = f(\text{DI}_{i})$. In panel (A) of Figure~\ref{Scheme} we have the APD restitution curve fitted to the kinetics of the Beeler-Reuter model~\cite{hastings2000alternans}. The APD restitution curve can be more complex, being a function of the history of the stimulation pacing protocol due to memory~\cite{fenton1999memory,tolkacheva2003condition, fox2002period,cherry2004suppression} and of calcium concentrations in the cells~\cite{diaz2004sarcoplasmic,restrepo2009spatiotemporal}.

Some recent studies have used Machine Learning techniques to predict chaotic time series from simulated and experimental data~\cite{pathak2018model,ramadevi2022chaotic,shahi2022machine,shahi2022prediction}, however, to our knowledge they have not been used for classification of chaos and approximation of Lyapunov exponents.     
In this paper, we combine the use of the classical Logistic map information~\cite{may1976simple} to train Artificial Neural Networks (ANNs), and the Beeler-Reuter APD heart map model~\cite{hastings2000alternans, beeler1977reconstruction} to verify the validity and select the most suitable Artificial Neural Network. With these elements we build an algorithm to detect chaos in experimental time series and we apply it to real-world data obtained from live frog hearts. An important point is to realize how the use of basic information data, i.e. the use of the basic and classical Logistic map as training data, combined with the a posteriori selection of the network using a heart map (in this case) allows to correctly detect the dynamics. This is a de facto proof of the universality
of the chaos dynamics information regardless of the problem being studied.

This paper is organized as follows. In Section~\ref{sec:2}, we describe the created DL algorithm to perform chaos analysis of heart time series (Subsections~\ref{subsec:21}-\ref{subsec:24} are devoted to explain in detail all the steps of the algorithm, and in Subsection~\ref{subsec:25} we provide the pseudocode). In Section~\ref{sec:3}, we apply such algorithm to experimental datasets obtained from frog heart dynamics (in Subsection~\ref{subsec:31} the approved animal protocol and experimental setup is described, in Subsection~\ref{subsec:32} we show the performance of the algorithm in the datasets and in Subsection~\ref{subsec:33} we carry out a brief comparison with standard techniques). Finally, in Section~\ref{sec:Conclusions} we draw some conclusions.

PyTorch~\cite{paszke2019pytorch} has been used to perform all the DL experiments in this work.

\section{DL Algorithm for Analyzing Chaotic Dynamics in Biological Time Series}\label{sec:2}
In this section we propose a new algorithm to analyze chaotic dynamics in biological (cardiac) time series. The algorithm consists on four steps:
\\

\textit{Step 1: DL Setup and Training.} $10$ randomly initialized recurrent-like Artificial Neural Networks with the architecture proposed in~\cite{ChaosDetCody23} are trained during $2,000$ epochs (early stopping and validation data are used) with time series of length $1,000$ from the Logistic Map (data creation as explained in~\cite{ChaosDetCody23, barrio2023deepdata}).
\\

\textit{Step 2: DL for Analyzing Chaotic Dynamics in a Mathematical Model.} A test analysis is performed using a mathematical model with the same nature as the experimental data (in our case, we consider an APD heart map~\cite{hastings2000alternans}) with each Artificial Neural Network trained in \textit{Step 1}.
\\

\begin{figure*}
	\begin{center}
		\includegraphics[width=0.69\textwidth]{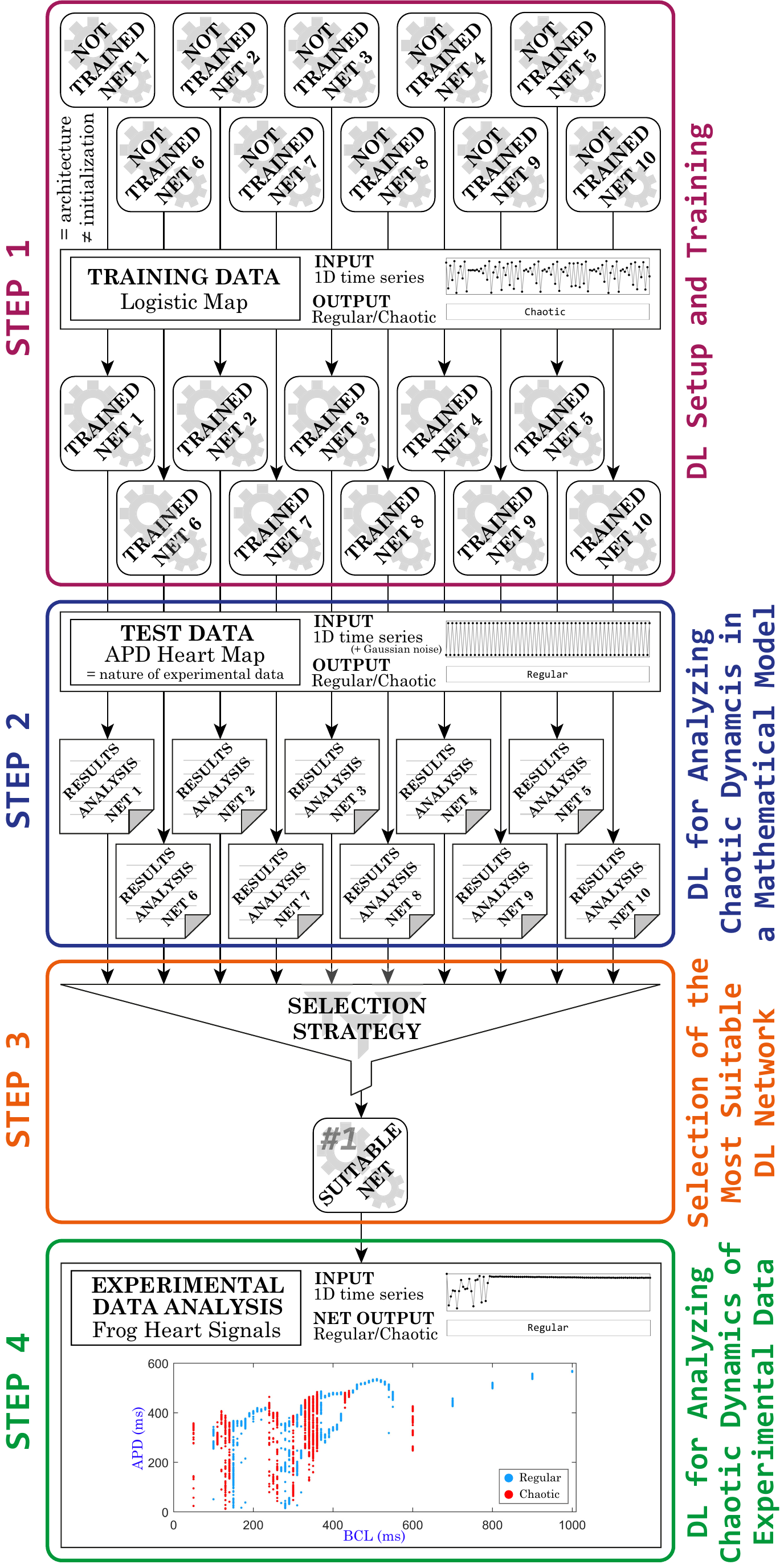}
		\caption{Schematic diagram of the algorithm for analyzing chaotic dynamics in experimental time series.}
		\label{Framework}
	\end{center}
\end{figure*}

\textit{Step 3: Selection of the Most Suitable DL Network.} Some criteria are applied on the results of \textit{Step~2} to detect automatically which is the most suitable network to perform the analysis of chaotic dynamics in biological (heart) time series.
\\

\textit{Step 4: DL for Analyzing Chaotic Dynamics of Experimental Data.} Analysis of chaotic dynamics is performed with the network chosen in \textit{Step 3} in the experimental data (in our case, APD time series from frog heart signals).
\\

In Subsections~\ref{subsec:21}-\ref{subsec:24}, all the steps of the algorithm are explained in detail. In subsection~\ref{subsec:25}, the pseudocode of the algorithm can be found. A complete schematic diagram of the four steps is depicted in Figure~\ref{Framework}.

\subsection{Step 1: DL Setup and Training}
\label{subsec:21}

Deep Learning (DL) is the branch of Machine Learning that uses Artificial Neural Networks (ANNs) to learn from data with different levels of abstraction. Since the introduction of ANNs, several architectures have been proposed and widely used for several applications~\cite{shahi2022prediction, hochreiter1997long, redmon2016you, liu2019arrhythmias}. One type is the Recurrent Neural Network (RNN) usually applied for sequential processing, which makes it an ideal choice when working with time series. Basic RNNs present several training problems as exploding/vanishing gradient and catastrophic forgetting. On the one hand, some authors have recently proposed alternative training algorithms based on dynamical systems theory to avoid gradient drawbacks~\cite{engelken2024gradient}. On the other hand, to try to alleviate those problems, new architectures as Long Short-Term Memory (LSTM) cells~\cite{hochreiter1997long} or Gated Recurrent Units (GRUs)~\cite{cho2014learning} have been developed. For our purpose we will use an architecture based on stacked LSTM cells with a final classification layer.

\begin{figure}[h!]
	\begin{center}
		\includegraphics[width = 0.48\textwidth]{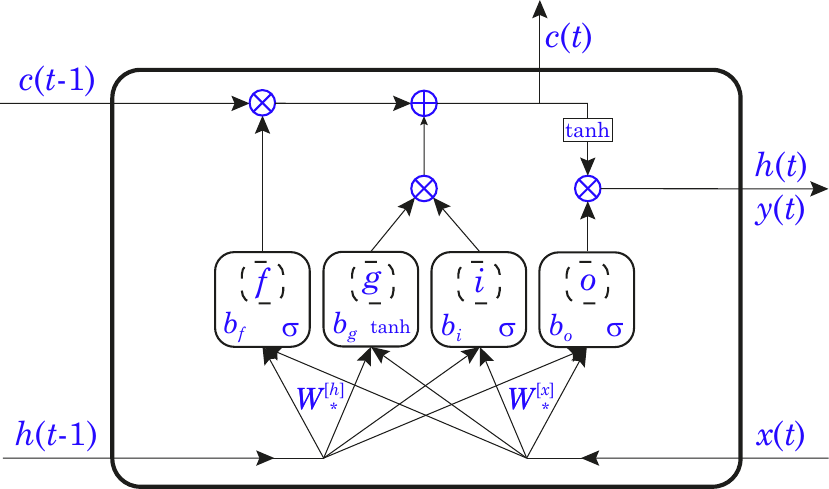}
		\caption{Scheme of an LSTM cell.}
		\label{LSTM}
	\end{center}
\end{figure}

A Long Short-Term Memory cell (Figure~\ref{LSTM}) is a DL architecture that processes information over time. At time step $t$, its inputs are an external data point $x(t)$ (usual input data for an ANN), a cell state $c(t-1)$, and a hidden state $h(t-1)$; and the outputs are the updated cell state $c(t)$, the updated hidden state $h(t)$, and the usual output $y(t)$ of an ANN (we take it equal to $h(t)$). The states $c(\cdot)$ and $h(\cdot)$ are devoted to keep information from previous time steps and are updated according to
\begin{equation*}
	\begin{array}{rcl}
		c(t) & = & f(t)\otimes c(t-1) + i(t)\otimes g(t),\\
		h(t) & = & o(t)\otimes\tanh(c(t)),
	\end{array}
\end{equation*}
where $\otimes$ is the element-wise product, and $f(t)$, $g(t)$, $i(t)$ and $o(t)$ are given by
\begin{equation*}
	\begin{array}{rcl}
		f(t) & = & \sigma(W_{f}^{[x]}x(t) + W_{f}^{[h]}h(t-1)+b_{f}),\\
		g(t) & = & \tanh(W_{g}^{[x]}x(t) + W_{g}^{[h]}h(t-1)+b_{g}),\\
		i(t) & = & \sigma(W_{i}^{[x]}x(t) + W_{i}^{[h]}h(t-1)+b_{i}),\\
		o(t) & = & \sigma(W_{o}^{[x]}x(t) + W_{o}^{[h]}h(t-1)+b_{o}).
	\end{array}
\end{equation*}
In these previous expressions, $\sigma(z)=1/(1+\text{exp}(-z))$ and $\tanh(z)={2\sigma(2z)-1}$ are the activation functions (sigmoid and hyperbolic tangent, respectively), and $W_{\ast}^{[\{x, h\}]}$ and $b_{\ast}$ ($\ast\;\in\{f, g, i, o\}$) are the trainable parameters (weights and biases, respectively) of the network. Note that because of the application of the sigmoid activation function, $f$, $i$ and $o$ act as gates that screen the information and memory of the network.

\begin{figure*}[h!]
	\begin{center}
		\includegraphics[width = \textwidth]{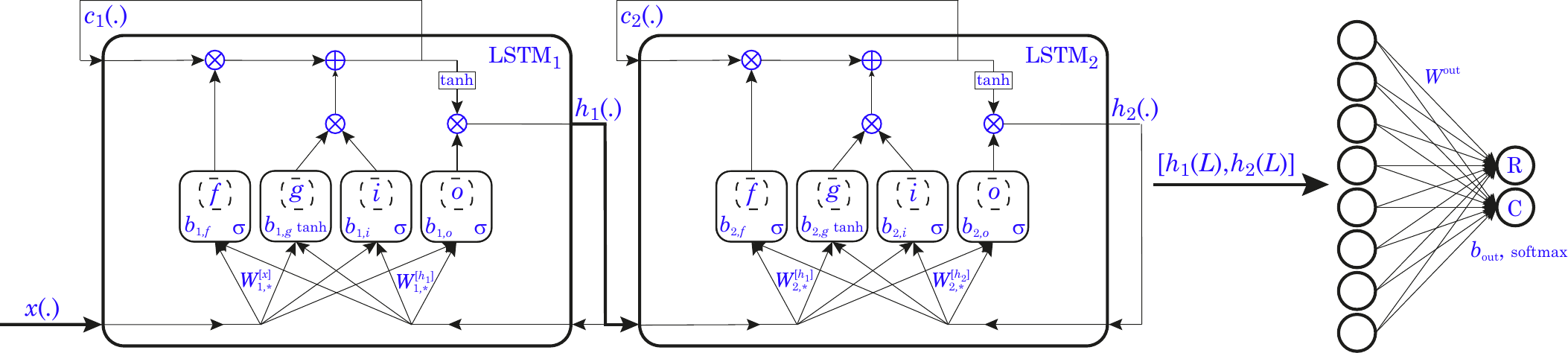}
		\caption{Graphical representation of the DL architecture used in this work ($L$ is the length of the input time series, R corresponds to regular class, and C to chaotic one). See the text for more details.}
		\label{Scheme_Net}
	\end{center}
\end{figure*}

The ANN that we use for chaos analysis (introduced in~\cite{ChaosDetCody23}) has two stacked trainable LSTM cells (both are unidirectional with bias term and states of dimension $4$) followed by a trainable linear classification layer of two neurons (one for each class: regular and chaotic) whose input is the last hidden state of both LSTM cells. Finally the softmax activation function, given by $\text{softmax}_i({z})=\text{exp}(z_i)/(\sum_{j=1}^{n}\text{exp}(z_j))$ with ${z}\in\mathbb{R}^{n}$, is applied to the output of the classification layer (to transform output values to scores, as usual for classification DL tasks). A graphical representation of the DL architecture used for chaos detection in this work is depicted in Figure~\ref{Scheme_Net}.

To fit the aforementioned trainable parameters (weights and biases), the network has to be trained using data (in our case, from the Logistic map). In fact, the training process consists in finding the value of the trainable parameters that minimizes the loss function (that measures the discrepancy between the network output and the expected output or label) for the given training dataset. As working with a classification task, we consider the standard cross-entropy loss function, and we use $L^{2}$-regularization (weight decay $\beta=10^{-5}$) to avoid overfitting. Therefore, the function to minimize is
\begin{equation*}
	{-\sum_{j=1}^{B}\log\left(\text{NN}(x^{j})\right)+\beta\sum_{k=1}^{K}w_k^{2},}
\end{equation*}
where $\text{NN}(x^{j})$ is the network output (once the softmax function has been applied) for input $x^{j}$, $K$ is the number of trainable parameters, $w_k$ represents a trainable parameter (a weight or a bias), and $B$ is the batch size (for optimization purposes, the training dataset is divided into disjoint subsets known as batches). The applied optimizer is Adam~\cite{kingma2014adam} with learning rate $0.008$. The network is trained for $2,000$ epochs using early stopping technique (that is, the final network has the value of the trainable parameters that provides the lower value for the validation dataset, a set of samples not present in the training set, during training process).

A key point related to the training process of the ANN is the selection of the training data. In our case, as the experimental data consists of short time series, we try to use basic and generic 
discrete data obtained from one of the most well-known discrete maps, the Logistic map~\cite{may1976simple}. One interesting fact in dynamical systems theory is the universality of numerous dynamical phenomena, like Feigenbaum constants~\cite{Feigenbaum78}, chaos dynamics and so on. Therefore, as training data we use generic data to not particularize too much the training process.

The Logistic map~\cite{may1976simple} is a well-known discrete one-dimensional dynamical system that presents great dynamical richness (period-doubling bifurcation, chaos, \dots) despite its simplicity. It is given by
\begin{equation}
	x_{i+1} = \alpha\,x_{i}\,(1-x_{i}),
	\label{LM_eq}
\end{equation}
where $x_{i}$ is the variable at the $i$-th iteration, and $\alpha$ is the bifurcation parameter (that can be interpreted as the growth rate of the system). The Logistic map is constructed in such a way that the variable only takes values in the interval $[0, 1]$. As it is a one-dimensional map, the Lyapunov exponent (LE)~\cite{argyris1994exploration} can be easily computed applying equation
\begin{equation}
	\text{LE} = \displaystyle\dfrac{1}{T}\sum_{j=1}^{T}\log\vert \alpha\,(1-2x_{j})\vert,
	\label{LM_LE}
\end{equation}
where the expression inside the absolute value corresponds to the derivative of the right-side of Equation~(\ref{LM_eq}) with respect to the system variable $x$, $\log$ is the natural logarithm, and $T$ corresponds to a large enough value of the iteration number. An appropriate number of transient iterations has to be computed prior to applying the formula.

\begin{figure}[h!]
	\begin{center}
		\includegraphics[width = 0.45\textwidth]{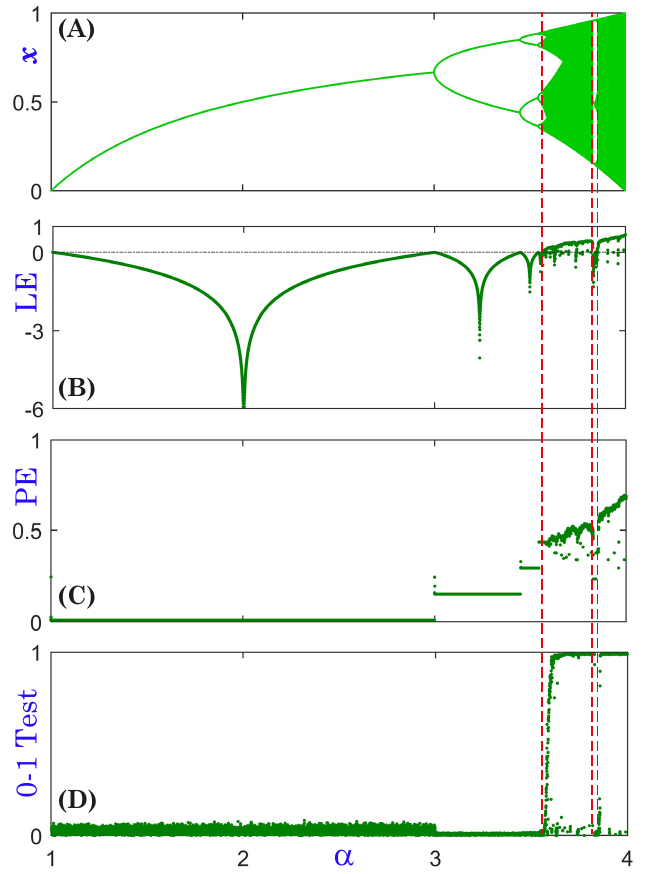}
		\caption{$\alpha$-parametric line of the Logistic map ($x_{0} = 0.5$). (A) Bifurcation diagram. (B) Lyapunov exponents. (C) Permutation Entropy (PE). (D) 0\,-1 test for chaos. The red vertical dashed lines across panels (A)-(D) indicate the boundary between regular and chaotic regions to show that the results provided by all techniques are consistent.}
		\label{LM_image}
	\end{center}
\end{figure}

In panel (A) of Figure~\ref{LM_image} we have represented the bifurcation diagram of the Logistic map when $\alpha\in[0,4]$ and $x_{0} = 0.5$. In this bifurcation diagram the dynamics of the Logistic map can be seen clearly: For small values of $\alpha$ the dynamics converge into an equilibrium point, later there is a cascade of period-doubling bifurcations which result in chaotic dynamics for values of $\alpha$ larger than $3.5$ approximately. In panel~(B), LEs have been depicted. We can check that for regular behavior (equilibrium points and periodic orbits) the LE is negative or zero, and it is positive for chaos. For comparison purposes, we also use other two chaos detection methods that have been used for time series classification in literature. In panel (C), we have depicted the (normalized) results obtained by applying the Permutation Entropy (PE) method (AntroPy~\cite{vallat2025antropy}, an open-source Python package, has been applied). Notice that to apply the Permutation Entropy technique~\cite{BandtPompe02, Riedl13}, we have to set the permutation order $n$ and the delay $\tau$. To that goal, we follow the parameter selection procedure of~\cite{Riedl13}: As we are working with a one-dimensional discrete dynamical system and a time series length of $1,000$, we fix the permutation order to $n=5$ (maximum $n$ value such that the length of the time series is greater than $5n!$) and the delay $\tau=1$. For the Permutation Entropy, lower values indicate less complexity, hence more regular behavior, whereas higher values indicate greater complexity, associated with more chaotic or irregular dynamics. In panel (D), we have represented the results provided by the 0\,-1 test for chaos (MATLAB implementation in~\cite{matthews2025chaos} has been used), where values around $0$ indicate regular behavior, and those that are approximately $1$ mean chaotic behavior. By comparing plots (B), (C), and (D), where different complexity measures have been applied, we observe that the results are consistent across them (see red vertical dashed lines that indicate the boundary between regular and chaotic regions).

To train the network with early stopping technique, the data is divided into three datasets: Training dataset contains data to learn from, validation dataset prevents overfitting, and test set checks the performance of the trained network. Each dataset contains time series and their corresponding LE computed with the classical technique in~\cite{argyris1994exploration} (that is used as label to check the behavior detected by the ANN). To obtain the time series, $12,000$ time steps are computed with Equation~\eqref{LM_eq} and the last $1,000$ correspond to the time series. To compute the LE, $12,000$ time steps are obtained with Equation~\eqref{LM_eq}, first $1,000$ points are discarded as transient, and Equation~\eqref{LM_LE} is applied with the remaining ones.

To obtain the training dataset, we create two raw datasets, one with initial condition $x_{0} = 0.5$ and other with $x_{0} = 0.9$. In both sets, the time series have length $1,000$ and the parameter $\alpha$ takes $12,000$ equidistant values in $[0,4)$. Data is screened deleting similar samples (two time series are similar if their distance in infinity norm is less than $10^{-4}$), and finally $2,000$ regular and $2,000$ chaotic samples are selected randomly to build such training dataset. For validation set, a raw dataset with time series of length $1,000$, initial condition $x_{0}=0.75$, and the parameter $\alpha$ taking $12,000$ equidistant values in $[0,4)$ is created. It is screened and $1,000$ time series of each dynamical behavior are selected randomly to obtain the final validation dataset. The test set is created similarly to validation dataset but taking $x_{0} = 0.8$ as the initial condition.

As shown in~\cite{ChaosDetCody23}, if we train $10$ randomly initialized ANNs with the aforementioned architecture and the built train and validation sets (trained networks are checked with the test set), a powerful tool to detect chaos in the Logistic map is obtained (accuracy greater than $95\%$ in all datasets and in all the trained networks).

All the parameters involved in the implementation of the network architecture and its training process are summarized in Table~\ref{Table_Parameters}.

\begin{Remark}
	Equivalent experiments have been carried out for different values of the time series length and the total number of epochs during training, but values $1,000$ and $2,000$, respectively, seem to give the best results. Notice that in~\cite{ChaosDetCody23}, length $1,000$ was also proposed as the best option to train networks for a chaos detection task.
\end{Remark}

\renewcommand{\arraystretch}{1.2}

\begin{table}[h!]
	\centering
	\resizebox{.46\textwidth}{!}{
	\begin{tabular}{|l|}
		\hline
		\cellcolor{gray!60}\textbf{DL Architecture}\\
		\hline
		\cellcolor{gray!20}\textbf{LSTM cell + LSTM cell + Classification layer} \\
		\hline
		LSTM \\
		\quad\quad Input size = 1 \\
		\quad\quad Hidden size = 4\\
		\quad\quad Number of layers = 2 (two stacked LSTM cells)\\
		\quad\quad Bias = True \\
		\quad\quad Bidirectional = False \\
		Classification (linear) layer\\
		\quad\quad Input size = 8 (last hidden state of both LSTM cells)\\
		\quad\quad Output size = 2 (2 classes: regular, chaotic)\\
		\quad\quad Activation function: Softmax\\
		\hline
		\hline
		\cellcolor{gray!60}{\centering\textbf{Loss Function}}\\
		\hline
		\cellcolor{gray!20}\textbf{Cross Entropy Loss + $\boldsymbol{L^{2}}$-regularization} \\
		\hline
		\quad\quad Weight decay ($\beta$) $= 10^{-5}$ \\
		\hline
		\hline
		\cellcolor{gray!60}{\textbf{Optimizer}}\\
		\hline
		\cellcolor{gray!20}\textbf{Adam optimizer} \\
		\hline
		\quad\quad Learning rate = $0.008$ \\
		\hline
		\hline
		\hline
		\cellcolor{gray!60}{\textbf{Training}}\\
		\hline
		\quad\quad 2,000 epochs + Early stopping \\
		\hline
		\cellcolor{gray!20}\textbf{Training data. Logistic map: $\boldsymbol{x_{i+1}=\alpha\,x_i\,(1-x_i)}$} \\
		\hline
		\quad\quad $x_0\in\{0.5,0.9\}$, $\alpha\in[0,4)$ \\
		\quad\quad Time series length = $1,000$\\
		\quad\quad $2,000$ regular samples + $2,000$ chaotic samples\\
		\hline
		\cellcolor{gray!20}\textbf{Validation data. Logistic map: $\boldsymbol{x_{i+1}=\alpha\,x_i\,(1-x_i)}$} \\
		\hline
		\quad\quad $x_0=0.75$, $\alpha\in[0,4)$ \\
		\quad\quad Length time series = $1,000$\\
		\quad\quad $1,000$ regular samples + $1,000$ chaotic samples\\
		\hline
	\end{tabular}}
	\caption{Summary of the parameters related to the network architecture and its training process.}
	\label{Table_Parameters}
\end{table}

\subsection{Step 2: DL for Analyzing Chaotic Dynamics in a Mathematical Model}
\label{subsec:22}

Our final objective (\textit{Step 4}) is to analyze chaotic dynamics of heart time series. We remark  that the $10$ ANNs have been trained with data from the Logistic map (\textit{Step 1}), not with heart-like data. Moreover, the time series that will be analyzed in \textit{Step~4} correspond to experimental recordings, that in general are short and noisy. Therefore, in this step we use some heart dynamics information, a simple cardiac map model, to perform several numerical tests on all trained ANNs. This is an important point as it connects generic and universal dynamics of the classical Logistic map with a more specific cardiac map model. That is, the training is in some way universal, but the selection of the particular ANN is done by a more specific mathematical model focused on the nature of the experimental data.

The APD restitution curve (which describes the dynamics of a single cardiac cell) fitted to the kinetics of Beeler-Reuter model~\cite{beeler1977reconstruction} gives rise to this discrete equation~\cite{hastings2000alternans} (named as Beeler-Reuter APD heart map in what follows):
\begin{equation*}
	\begin{array}{c}
		\vspace{0.2cm}\text{APD}_{i+1} = 258 + 125\exp\big({-0.068(\text{DI}_{i}-43.54)}\big)\\
		- 350\exp\big({-0.028(\text{DI}_{i}-43.54)}\big),
	\end{array}
\end{equation*}
where $\text{APD}_{i+1}$ is the Action Potential Duration (APD, duration from the onset of cell depolarization to the completion of repolarization) of the $(i+1)$-th stimulus and $\text{DI}_{i} = n\,\text{BCL}-\text{APD}_{i}$ is the Diastolic Interval (DI, recovery time of the cell) of previous stimulus. Therefore, the map equation can be rewritten as
\begin{equation}
	\begin{array}{c}
		\vspace{0.2cm}\text{APD}_{i+1} = 258 + 125\exp\big({-0.068(n\,\text{BCL}-\text{APD}_{i}-43.54)}\big)\\
		- 350\exp\big({-0.028(n\,\text{BCL}-\text{APD}_{i}-43.54)}\big),
	\end{array}
	\label{BR_eq}
\end{equation}
where $\text{BCL}$ is the Basic Cycle Length (BCL), that is, the time between two consecutive pacing stimuli, and $n$ is the parameter block (lower $n\in\mathbb{N}$ such that $n\,\text{BCL}-\text{APD}_{i}\geq \text{DI}_{\min}$, with $\text{DI}_{\min}$ the minimum DI whose value is set to $43.54$~ms). Notice that the BCL can be considered as the bifurcation parameter as it is independent of discrete time.

Since the Beeler-Reuter APD heart map is a one-dimensional map, the Lyapunov exponent~\cite{argyris1994exploration} can be obtained by applying the formula
\begin{equation}
	\begin{array}{c}
		\vspace{0.2cm}\text{LE} = \displaystyle\dfrac{1}{T}\sum_{j=1}^{T}\log \big\vert 8.5\exp\big({-0.068(n\,\text{BCL}-\text{APD}_{j}-43.54)}\big)\\
		- 9.8\exp\big({-0.028(n\,\text{BCL}-\text{APD}_{j}-43.54)}\big)\big\vert,
	\end{array}
	\label{BR_LE}
\end{equation}
where the expression inside the absolute value corresponds to the derivative of the right-side of Equation~(\ref{BR_eq}) with respect to the system variable APD, $\log$ is the natural logarithm, and $T$ corresponds to a large enough value of the iteration number. An appropriate number of transient iterations must be calculated before applying the formula.

\begin{figure}[h!]
	\begin{center}
		\includegraphics[width = 0.45\textwidth]{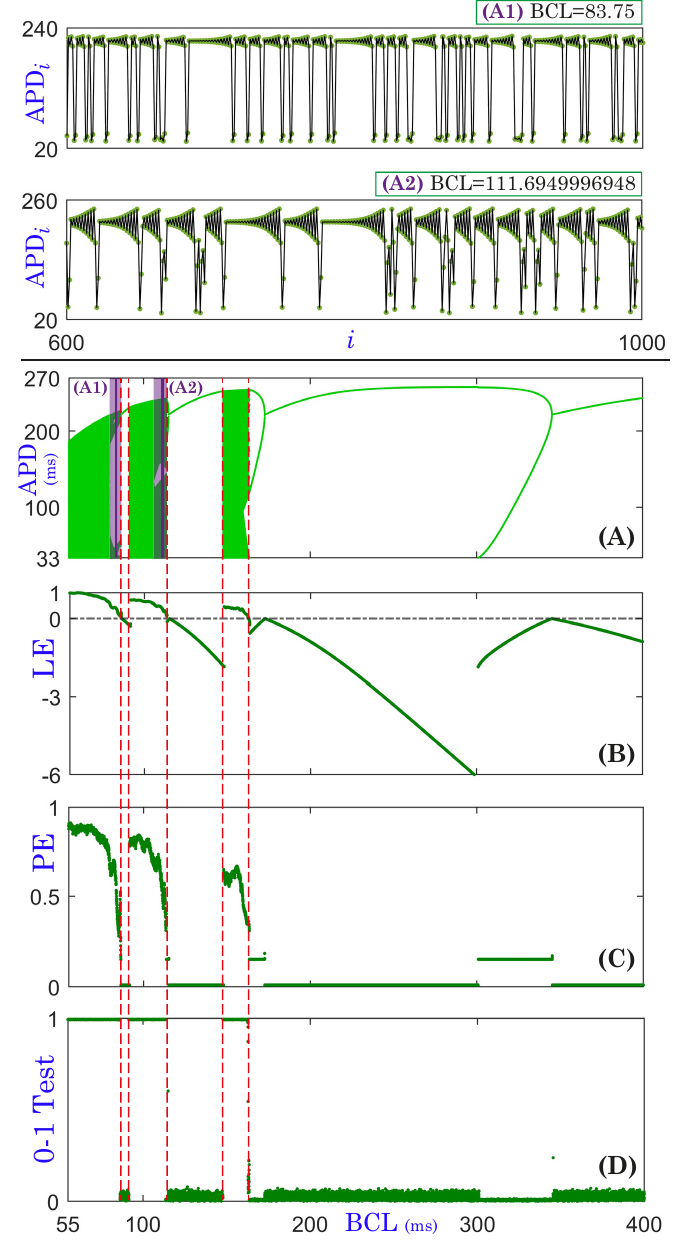}
		\caption{$\text{BCL}$-parametric analysis of the Beeler-Reuter APD heart map ($\text{APD}_{0} = 240\text{ ms}$). (A) Bifurcation diagram. Purple shading corresponds to regions where the network fails the most in the classification between regular and chaotic dynamics. (A1) and (A2) are two of these failing time series (in green we have the points of the time series, and we have joined such points with black segments for the ease of viewing). (B) Lyapunov exponents (LEs). (C) Permutation Entropy (PE). (D) 0\,-1 test for chaos. The red vertical dashed lines across panels (A)-(D) indicate the boundary between regular and chaotic regions to show that the results provided by all techniques are consistent.}
		\label{BR_image}
	\end{center}
\end{figure}

In panel (A) of Figure~\ref{BR_image}, we have depicted the bifurcation diagram of the Beeler-Reuter APD heart map for $\text{BCL}\in[55,400)\text{ ms}$ and $\text{APD}_{0} = 240\text{ ms}$. If we move from higher to lower BCL values in such a bifurcation diagram we can see equilibrium points, a period-doubling bifurcation, and a $2$:$1$ block (the cell responds one of each two pacing stimuli) followed by a period-doubling, chaos, and higher-order blocks (see~\cite{hastings2000alternans} for more dynamical information). In panel (B), the LEs calculated with Equation~(\ref{BR_LE}) have been represented. In this panel it can be seen that regular behavior (equilibrium points and periodic orbits) corresponds to negative or zero LEs, and chaos to positive LE values. In panel (C), we have depicted the (normalized) results obtained applying Permutation Entropy (AntroPy~\cite{vallat2025antropy}, an open-source Python package, has been applied). As indicated for the Logistic map, to apply the Permutation Entropy technique we have to set the permutation order and the delay. As we are also working with a one-dimensional discrete dynamical system and a time series length of $1,000$, as clarified before, we consider permutation order $5$ and delay $1$. For this technique, lower values indicate less complexity, hence more regular behavior, whereas higher values indicate greater complexity, associated with more chaotic or irregular dynamics. In panel (D), we have represented the results provided by the 0\,-1 test for chaos (MATLAB implementation in~\cite{matthews2025chaos} has been used), where values around $0$ indicate regular behavior, and those that are approximately $1$ mean chaotic behavior. By comparing plots (B), (C), and (D), where different complexity measures have been applied, we observe that the results are consistent across them (see red vertical dashed lines that indicate the boundary between regular and chaotic regions).

At this point, we are going to perform a chaos analysis of one BCL-parametric line (see Figure~\ref{BR_image}) of this APD heart map using the networks trained in \textit{Step $1$}. Such BCL-parametric line has been obtained taking $12,000$ equidistant values for the BCL in the interval $[55, 400)\text{ ms}$ and initial condition $\text{APD}_{0}=240\text{ ms}$. Equation~(\ref{BR_eq}) is applied for $20,500$ time points. First $500$ points are discarded as transient, and the remaining ones are used for the LEs computation (Equation~(\ref{BR_LE})). The time series obtained for each BCL value are constructed with the last $1,000$ time points. Later, a component-wise Gaussian random noise with strength $\mu\in\{0, 0.5, 1.0\}$ is added to each time series (notice that when $\mu = 0$, no noise is added). The analyses of time series with Gaussian noise are repeated $100$ times to give results as mean$\pm$standard deviation. Notice that the time series from the Logistic map take values in $[0, 1]$. This does not occur with this Beeler-Reuter APD heart map, so we normalize the data to the interval $[0, 1]$ before performing the DL analysis of chaotic dynamics (a random number sampled uniformly in such interval is assigned to constant time series and a linearly mapping between the range of non-constant samples to the interval $[0, 1]$ is done otherwise).

\begin{Remark}
	The initial condition and the times used for LEs computation in the APD heart map are taken as in~\cite{lewis1990chaotic}.
\end{Remark}

Now, we present different accuracy indicators (adapted to the use of experimental data and the cardiac model) to provide information to  select  in \textit{Step~3} the most appropriate network (among all computed in \textit{Step~1}) to analyze heart data under noise and the absence of it.

The \textsl{Accuracy} (percentage of samples that have been correctly classified) is the measure proposed to check the performance of the networks. It is defined as
\begin{equation*}
	\text{Accuracy }(\%) = \dfrac{T_{R}+T_{C}}{T_{R}+T_{C}+F_{R}+F_{C}}\cdot 100,
\end{equation*}
where $T_{R}$ and $T_{C}$ are the number of true regular and true chaotic samples (that is, the number of samples that are correctly classified by the network as regular and chaotic, respectively); and $F_{R}$ and $F_{C}$ are the false regular and false chaotic samples (that is, the number of samples that are incorrectly classified by the DL network as regular and chaotic, respectively). Notice that if the dataset in which this formula is applied is unbalanced (we do not have the same number of samples of each dynamical behavior) as occurs in our BCL-parametric line ($19.708\%$ of the samples are chaotic according to the LE value, and the remaining $80.292\%$ have a regular behavior), this measure cannot be reliable. For example, in a dataset with a large number of regular and a small number of chaotic samples, even if all the chaotic samples are not detected correctly, the accuracy can be close to $100\%$ if all the regular samples are correctly classified and, this does not mean that the network is able to perform the chaos detection analysis properly. To avoid this problem, we use other variants of the accuracy, the \textsl{Accuracy Chaotic} and \textsl{Accuracy Regular} that compute the percentage of chaotic and regular samples, respectively, that are correctly classified. The corresponding formulas are as follows
\begin{equation*}
	{\text{Accuracy Chaotic }}(\%) = \dfrac{T_{C}}{T_{C}+F_{R}}\cdot 100,
\end{equation*}
\begin{equation*}
	{\text{Accuracy Regular }}(\%) = \dfrac{T_{R}}{T_{R}+F_{C}}\cdot 100.
\end{equation*}

\begin{Remark}
	If we consider chaotic as the positive class and regular as the negative one, accuracy chaotic corresponds to the sensitivity or recall (in $\%$), and accuracy regular is the specificity (in $\%$).
\end{Remark}

Notice that a regular dynamical behavior includes mainly two different types of dynamics: equilibrium points (EPs) and periodic orbits (POs). To complete our analysis we also compute the \textsl{Accuracy EPs} and \textsl{Accuracy POs}, that is, the percentage of equilibrium points and periodic orbits, respectively, that have been classified correctly as regular:
\begin{equation*}
	{\text{Accuracy EPs }}(\%) = \dfrac{T_{\text{R\,|\,EPs}}}{T_{\text{R\,|\,EPs}}+F_\text{C\,|\,EPs}}\cdot 100,
\end{equation*}
\begin{equation*}
	{\text{Accuracy POs }}(\%) = \dfrac{T_{\text{R\,|\,POs}}}{T_{\text{R\,|\,POs}}+F_\text{C\,|\,POs}}\cdot 100,
\end{equation*}
with $T_{\text{R\,|\,EPs}}$ and $T_{R\,|\,\text{POs}}$ the number of equilibrium points and periodic orbits, respectively, that have been classified correctly as regular; and $F_{\text{C\,|\,EPs}}$ and $F_{\text{C\,|\,POs}}$ the number of equilibrium points and periodic orbits, respectively, that are incorrectly classified as chaotic. In our parametric line, $79.803\%$ of the regular samples are equilibrium points and the remaining $20.197\%$ present periodic behavior, so we consider that the measures that we have just defined are important to check that with DL both dynamics are detected properly as regular.

Finally, we define an indicator to detect if the network detection is robust against noise. We refer to it as $\textit{diff}_{0-\tilde{\mu}}$ and it corresponds to the percentage of samples that the network does not detect with the same behavior in the analysis without noise ($\mu = 0$) and with noise strength $\tilde{\mu}\neq0$ (in our analysis $\tilde{\mu} = 0.5$ or $\tilde{\mu} = 1.0$). That is, if 
$${\footnotesize{S_{\mu} = \{\text{classification obtained from data with noise strength }\mu\},}}$$ then
\begin{equation*}
	\textit{diff}_{0-\tilde{\mu}} = 100 - \dfrac{\#[S_{\mu = 0}\cap S_{\mu = \tilde{\mu}}]}{\text{ total number of samples }} \cdot 100,
\end{equation*}
where $\#$ is used to refer to the cardinality.

\begin{Remark}
	The tests performed in this paper with Beeler-Reuter APD heart map can also be performed with Lewis and Guevara APD heart map~\cite{lewis1990chaotic} and similar/equivalent results are obtained.
\end{Remark}

\subsection{Step 3: Selection of the Most Suitable DL Network}
\label{subsec:23}

Once we have performed the DL chaos analysis of the APD heart map (\textit{Step 2}) we have to apply some criteria to choose automatically (without direct human supervision) a robust network against noise that can perform properly the chaos detection in the Beeler-Reuter APD heart map. We expect that the selected network will be able to perform properly the DL chaos analysis of biological time series of frog heart signals.

The automatic selection criteria takes into account the accuracy-like measures and the indicator $\textit{diff}_{0-\tilde{\mu}}$ (for $\tilde{\mu}\in\{0.5, 1.0\}$) defined in \textit{Step 2}. In particular, our selection criteria is as follows: All accuracies (accuracy, accuracy chaotic, accuracy regular, accuracy EPs and accuracy POs) have to be greater than $75\%$ in mean (to ensure good performance in chaos detection task), and $\textit{diff}_{0-0.5}$ and $\textit{diff}_{0-1.0}$ have to be less than $1\%$ in mean (to ensure robustness in the detection against noise). Moreover, we impose that the standard deviation for all the measures and indicators has to be less than $1\%$ (to ensure that results are quite independent of randomness).

Notice that we are checking if any of the $10$ trained networks (\textit{Step 1}) has gone further, and besides having learned to detect chaos in the Logistic map, it is able to generalize its detection to discrete heart dynamics. Note that the ANNs have only been trained with dynamics from the Logistic map, the data was not preprocessed in any special way, and no extra dynamical information has been given to the networks, so the proposed task is not easy and we consider that the aforementioned criteria is reasonable.

\begin{table*}[htb]
	\centering
	\resizebox{.658\textwidth}{!}{
	\begin{tabular}{|p{4cm}||p{2.3cm}|p{2.3cm}|p{2.3cm}|}
		\hline
		& \cellcolor{gray!60}{$\boldsymbol{\mu = 0}$} & \cellcolor{gray!60}{$\boldsymbol{\mu = 0.5}$} & \cellcolor{gray!60}{$\boldsymbol{\mu = 1.0}$} \\
		\hline
		\hline
		\cellcolor{gray!20}{\rm{\textbf{Accuracy (\%)}}} & $95.092$ & $94.752\pm0.050$ & $94.651\pm0.060$ \\
		\hline
		\cellcolor{gray!20}{\rm{\textbf{Accuracy Chaotic (\%)}}} & $76.195$ & $75.951\pm0.071$ & $75.552\pm0.112$ \\
		\hline
		\cellcolor{gray!20}{\rm{\textbf{Accuracy Regular (\%)}}} & $99.730$ & $99.367\pm0.060$ & $99.339\pm0.069$ \\
		\hline
		\cellcolor{gray!20}{\rm{\textbf{Accuracy EPs (\%)}}} & $100$ & $99.545\pm0.075$ & $99.542\pm0.087$ \\
		\hline
		\cellcolor{gray!20}{\rm{\textbf{Accuracy POs (\%)}}} & $98.666$ & $98.664\pm0.009$ & $98.541\pm0.037$ \\
		\hline
		\hline
		\cellcolor{gray!20}{\rm{$\boldsymbol{\textbf{\textit{diff}}_{0-\tilde{\mu}}}$ \textbf{(\%)}}} & - & $0.361\pm0.049$ & $0.493\pm0.059$\\
		\hline
	\end{tabular}}
	\caption{Results of the DL chaos analysis of the Beeler-Reuter APD heart map with the network chosen by the criteria applied in \textit{Step~3}. Notice that $\mu = 0$ corresponds to the analysis without noise. Results for $\mu\in\{0.5, 1.0\}$ are given as mean$\pm$standard deviation for $100$ trials.}
	\label{Table_BR}
\end{table*}

In Table~\ref{Table_BR} we have the results for the DL analysis of chaotic dynamics (\textit{Step 2}) of the network that satisfies all the imposed criteria. In particular, for the cases with noise (second and third columns), the analyses have been performed $100$ times because of randomness, and results correspond to mean$\pm$standard deviation.

Let us focus on the results of the without noise case ($\mu = 0$). Notice that the value of all the accuracies (except Accuracy Chaotic) are greater than $95\%$. Notice that both types of regular behavior (equilibrium points and periodic orbits) are properly detected in almost all the cases. It is necessary to check why the Accuracy Chaotic is lower than the other accuracy measures (but still greater than $75\%$). The majority of the samples that are incorrectly detected  as regular when their behavior is chaotic are in the boundary parts shaded in purple in the bifurcation diagram of the APD heart map in panel (A) of Figure~\ref{BR_image}. In panels (A1) and (A2) of this figure we have an example of a failing sample on each purple region ($\text{BCL}=83.75\text{ ms}$ and $\text{BCL}=111.6949996948\text{ ms}$, respectively). Notice that both samples are quite similar. As we can see, the general behavior is chaotic. However, it highlights the upper parts of the time series in which, compared with the general range of the APD values of the time series, there is not much variability. This is a behavior that we do not expect to be present in the Logistic map where the networks are trained. Taking into account that the data used to train the networks does not have extra dynamical information beyond that provided by the time series, and this behavior is not present, to require the network to detect it is very demanding.

For the analyses with noise (noise strength $\mu\in\{0.5, 1.0\}$), all the accuracy results are quite good and similar to those given by the without noise case (with a standard deviation less than $0.15$ in all cases). This is confirmed by the $\textit{diff}_{0-0.5}$ and $\textit{diff}_{0-1.0}$ indicators that are lower than $0.5\%$ in mean and with standard deviation lower than $0.06$, increasing its value with noise strength as expected.

\subsection{Step 4: DL for Analyzing Chaotic Dynamics of Experimental Data}
\label{subsec:24}

So far, we have trained $10$ randomly initialized recurrent-like Artificial Neural Networks using data from the Logistic map (\textit{Step 1}), we have performed DL chaos analyses of an APD heart map with all the trained networks (\textit{Step 2}), and we have applied some criteria on these analyses to choose one network that performs well and is robust against noise (\textit{Step 3}). Now, we have all the ingredients to try a DL chaos analysis of biological time series from heart dynamics.

When training the networks, the data comes from the Logistic map. As already highlighted before, this equation is constructed in such a way that time series values are in the interval $[0,1]$. In general, the experimental data is not in the interval $[0,1]$, so we have to normalize it to that interval (the trained networks will not be able to process data properly in whatever other rank). With the data already normalized, time series are given as input to the chosen network in \textit{Step 3}. The output of the network will give us the information about the behavior (regular or chaotic) of such time series. Therefore, the defined algorithm allows performing chaos analysis of heart time series without human supervision.

\begin{Remark}
	Notice that the trained network is a recurrent-like neural network, therefore, the length of the input is not fixed and time series of whatever length can be processed. For example, with other architectures as the Multi-Layer Perceptron (obtained when perceptrons are stacked) the input size has to be constant.
\end{Remark}

\subsection{Pseudocode of the Algorithm for Analyzing Chaotic Dynamics in Biological Time Series}
\label{subsec:25}

\begin{algorithm*}[htb]
	\caption{DL Algorithm for Analyzing Chaotic Dynamics in Biological Time Series}\label{our_algorithm}
	\begin{algorithmic}[1]
		\Procedure{\textit{Step 1 }}{$\text{num\_nets}=10, \text{ANN}_\text{arch}, \text{dataLM}_{\text{train}}, \text{dataLM}_{\text{val}}, \text{dataLM}_{\text{test}}$}\Comment{ANNs Framework}
		\ParallelFor{$i = 1$ to $\text{num\_nets}$}
		\State $\text{Net}(i)\gets {\large{\texttt{RandomInitialization}}}(\text{ANN}_\text{arch})$\Comment{Random initialization of ANN trainable parameters}
		\State $\text{T-Net}(i)\gets {\large{\texttt{Train}}}(\text{Net}(i), \text{dataLM}_{\text{train}}, \text{dataLM}_{\text{val}})$\Comment{Training}
		\State ${\large{\texttt{Test}}}(\text{T-Net}(i), \text{dataLM}_{\text{test}})$\Comment{Test}
		\State \textbf{return} $\text{T-Net}(i)$
		\EndParallelFor
		\EndProcedure
		
		\Procedure{\textit{Step 2 }}{$\mu_1 = 0$, $\mu_2=0.5, \mu_3=1.0, \text{T-Net}(1),\cdots,\text{T-Net}(\text{num\_nets}), \text{dataBR}_{\text{BCL-line}}$}\Comment{Analysis APD Map}
		\ParallelFor{$i = 1$ to $\text{num\_nets}$}
		\For{$j = 1$ to $3$}
		\State $\text{dataBR}_{\text{BCL-line},\,{\mu}_j}\gets \text{dataBR}_{\text{BCL-line}}^{\text{time series}}+\mu_j\cdot Gaussian(0, 1)$\Comment{Add noise with strength $\mu_j$}
		\State $\text{N-dataBR}_{\text{BCL-line},\mu_j}\gets {\large{\texttt{normalization}}}(\text{dataBR}_{\text{BCL-line},\mu_j})$\Comment{Data normalization}
		\State $\text{result}_{\mu_j}(i)\gets {\large{\texttt{DLAnalysis}}}(\text{T-Net}(i), \text{N-dataBR}_{\text{BCL-line}, \mu_j})$\Comment{DL chaos analysis}
		\State $\text{acc-like}_{\mu_j}(i)\gets {\large{\texttt{AccuracylikeMeasures}}}(\text{result}_{\mu_{j}}(i), \text{N-dataBR}_{\text{BCL-line}}^{\text{labels}})$\Comment{Accuracy-like measures}
		\State $\textit{diff}_{\mu_{1}-\mu_{j}}(i)\gets {\large{\texttt{RobustnessIndicator}}}(\text{result}_{\mu_1}(i), \text{result}_{\mu_j}(i))$\Comment{Robustness indicator}
		\State \textbf{return} $\text{acc-like}_{\mu_j}(i), \textit{diff}_{\mu_1-\mu_j}(i)$
		\EndFor
		\EndParallelFor
		\EndProcedure
		
		\Procedure{\textit{Step 3 }}{$\text{acc-like}_{\mu_j}(i)$, $\textit{diff}_{[\mu_1=0]-\mu_j}(i)$ with $i=1,\cdots,\text{num\_nets}$, $j=1,2,3$ }\Comment{Selection of suitable ANN}
		\ParallelFor{$i = 1$ to $\text{num\_nets}$}
		\If{$\text{acc-like}_{\mu_1}(i)>75\%$ \textbf{and} $\text{acc-like}_{\mu_2}^{\text{mean}}(i)>75\%$
			\textbf{and} $\text{acc-like}_{\mu_2}^{\text{std}}(i)<1\%$
			\textbf{and} $\text{acc-like}_{\mu_3}^{\text{mean}}(i)>75\%$
			\textbf{and} $\text{acc-like}_{\mu_3}^{\text{std}}(i)<1\%$
			\textbf{and} $\textit{diff}_{\mu_1-\mu_2}^{\text{ mean}}(i)<1\%$ \textbf{and}
			$\textit{diff}_{\mu_1-\mu_2}^{\text{ std}}(i)<1\%$ \textbf{and} $\textit{diff}_{\mu_1-\mu_3}^{\text{ mean}}(i)<1\%$
			\textbf{and} $\textit{diff}_{\mu_1-\mu_3}^{\text{ std}}(i)<1\%$}
		\State $\text{suitableIndex}\gets i$\Comment{Index of the ANN that satisfies the criteria}
		\State \textbf{return} $\text{suitableIndex}$
		\EndIf
		\EndParallelFor
		\EndProcedure
		
		\Procedure{\textit{Step 4 }}{$\text{T-Net}(\text{suitableIndex}), \text{dataExperimental}^{\text{time series}}$}\Comment{Analysis Experimental Data}
		\ParallelFor{$k = 1$ to ${\large{\texttt{size}}}(\text{dataExperimental}^{\text{time series}})$}
		\State $\text{N-dataExperimental}\gets {\large{\texttt{normalization}}}(\text{dataExperimental}^{\text{time series}})$\Comment{Data normalization}
		\State $\text{result}\gets {\large{\texttt{DLAnalysis}}}(\text{T-Net}(\text{suitableIndex}), \text{N-dataExperimental})$\Comment{DL chaos analysis experimental data}
		\State \textbf{return} $\text{result}$
		\EndParallelFor
		\EndProcedure
	\end{algorithmic}
\end{algorithm*}

In Algorithm~\ref{our_algorithm} we have the pseudocode of the DL algorithm for analyzing chaotic dynamics in biological time series that we have described in Subsections~\ref{subsec:21}-\ref{subsec:24}. For simplicity and better comprehension, we have used some abbreviations. In particular, with $\texttt{num\_nets}$ we refer to the number of ANNs with the same architecture that are randomly initialized and trained (it is set to $10$ as indicated in Subsection~\ref{subsec:21}). $\texttt{ANN}_\texttt{arch}$ corresponds to the architecture of the network described in Subsection~\ref{subsec:21}. $\texttt{dataLM}_{\ast}$ with $\ast \; \in\{\texttt{train}, \texttt{val}, \texttt{test}\}$ is the data from the Logistic map used to train, validate, and test the network, respectively (such data includes the time series $\texttt{dataLM}_{\ast}^{\texttt{time series}}$ and the corresponding labels $\texttt{dataLM}_{\ast}^{\texttt{labels}}$). The other two datasets used in the algorithm are the $\text{BCL}$-parametric line of the Beeler-Reuter APD heart map referred as $\texttt{dataBR}_{\texttt{BCL-line}}$ (with $\texttt{dataBR}_{\texttt{BCL-line}}^{\texttt{time series}}$ the time series, and $\texttt{dataBR}_{\texttt{BCL-line}}^{\texttt{labels}}$ the labels), and the experimental dataset of heart dynamics indicated as $\texttt{dataExperimental}^{\texttt{time series}}$ (note that in this case we do not have the labels $\texttt{dataExperimental}^{\texttt{labels}}$). When using the notation $\texttt{acc-like}_{\mu_j}(i)$ for $j=1,2,3$ and $i=1,\cdots,\texttt{num\_nets}$ (line 15 of the code), we consider that we have a list that includes the accuracy, the accuracy chaotic, the accuracy regular, the accuracy EPs and the accuracy POs of net $i$ when noise with strength $\mu_j$ is added to the data. In addition, these values (as well as the values of the robustness indicator) are given as mean$\pm$standard deviation (mean$\pm$std). Therefore in the criteria of \textit{Step 3} (see line 23), for example, condition $\texttt{acc-like}_{\mu_j}^{\texttt{mean}}(i)>75\%$ indicates that all the elements of the list have to be greater than $75\%$ in mean.

\begin{Remark}
	Notice that in \textit{Step 2}, for case $j=1$, we have $\mu_j=\mu_1=0$, so in the part of the algorithm where Gaussian noise is added (line 12 of the code) the data does not change. Moreover, for this value of $j$ in this step of the algorithm, the computation of the robustness indicator $\textit{diff}_{\mu_1-\mu_j}=\textit{diff}_{\mu_1-\mu_1}$ (line 16 of the code) is trivial as it will always be equal to $0\%$ (for this reason, it is not included in the criteria of \textit{Step 3}, see line 23 of the code). An \texttt{if} statement could be added to avoid calculating it.
\end{Remark}

\section{Results of DL Algorithm for Analyzing Chaotic Dynamics in Biological Time Series: Frog Heart Signals}
\label{sec:3}
In this section we explain how the experimental data (frog heart signals) is obtained (Subsection~\ref{subsec:31}), we apply the previous detailed Algorithm~\ref{our_algorithm} to analyze its chaotic dynamics (Subsection~\ref{subsec:32}), and we briefly compare with the results provided by standard techniques (Subsection~\ref{subsec:33}).

\subsection{Data. Frog Cardiomyocyte Recordings}
\label{subsec:31}

In this subsection we explain in detail the protocol of the experiment and how it has been carried out to obtain the frog heart signals.

\begin{figure}[h!]
	\begin{center}
		\includegraphics[width=0.4\textwidth]{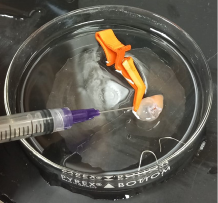}
		\caption{Photo taken during an experiment with a frog heart in the laboratory of Professor Flavio H. Fenton in Georgia Tech.}
		\label{Photo_Frog}
	\end{center}
\end{figure}

The frog heart experiments were performed following an approved Georgia Tech, Institutional Animal Care and Use Committee (IACUC) protocol $\#\, A100673$. Frogs were euthanized via fast decapitation following by Pithing of the brain and spinal cord which ensures the cessation of neural activity avoiding any pain. 
The heart is then quickly excised and cannulated via the aorta using a syringe filled with Tyrode solution, a blood substitute, to wash out all the blood (see Figure~\ref{Photo_Frog}). This process ensured a blood-cloths free preparation and allowed the heart to be kept physiologically viable in a container at room temperature, also filled with Tyrode solution, throughout the experiment. 

Transmembrane voltage signals were recorded as described in previous preparations~\cite{cherry2008visualization,cherry2007tale}. Briefly, individual heart cells from the whole heart were impaled with a fine-tipped micro-electrode, which was connected to a high-impedance amplifier to minimize signal loss. The amplified signals were routed to a data acquisition (DAQ) board interfaced with a computer for recording and analysis. This setup enabled precise voltage recordings of the electrical activity of the cardiac cells.

To induce variations in the Basic Cycle Length (BCL) of pacing, an isolated external current stimulation device was employed. The stimulating electrode was positioned about $0.5$ cm from the recording micro-electrode and delivered currents at twice the excitation threshold. This ensured effective and consistent stimulation of the cardiac tissue for the experimental protocols.

We consider two experimental datasets (Dataset 1 and Dataset 2) obtained as explained before from two different frog hearts. Both sets of samples contain voltage time series with several APDs, each one with a different BCL. That is, such time series correspond to the Action Potential Duration of frog heart dynamics for different pacing rates. All signals were recorded after the BCL was left for several minutes of accommodation to that BCL.

In Dataset 1 we have $52$ voltage time series whose minimum length (number of points) is $15$ and the maximum is $205$. In this case, $\text{BCL}\in[50, 1000]$ (ms). In panel (A) of Figure~\ref{ExpResults} (ignore colors for now) we have the bifurcation diagram obtained for this experimental dataset ($x$-axis is the pacing rate $\text{BCL}$ and $y$-axis is the APD-time series). In panel (B) of this figure (ignore colors for now) we have the length (number of points) of each time series.

Dataset 2 contains $24$ voltage time series. Except for an unusual long time series of $521$ points, the remaining ones have a minimum length (number of points) of $21$ and a maximum of $138$. For this dataset, $\text{BCL}\in[200, 1000]$ (ms). In panel (A) of Figure~\ref{ExpResults2} (ignore colors for now) we have the bifurcation diagram obtained for these experimental frog heart signals ($x$-axis is the pacing rate $\text{BCL}$ and $y$-axis is the APD-time series). In panel (B) of such figure (ignore colors for now) we have the length (number of points) of each time series.


\subsection{Results}
\label{subsec:32}

We are going to apply the DL algorithm for chaos analysis of heart time series defined in Section~\ref{sec:2} to some experimental data. That is, once \textit{Steps 1-3} (see Subsections~\ref{subsec:21}-\ref{subsec:23}) have been carried out, and a network that we expect to be suitable for chaos detection has been obtained, we use it in the experimental datasets of frog heart signals defined in Subsection~\ref{subsec:31}.

\begin{figure}
	\begin{center}
		\includegraphics[width=0.48\textwidth]{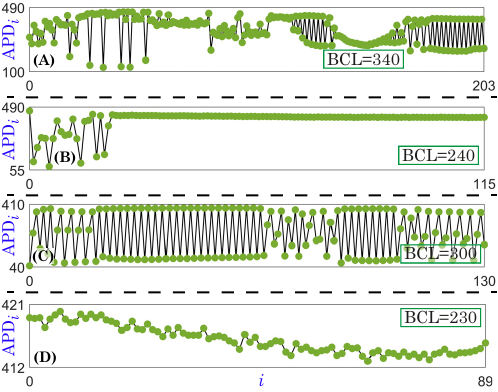}
		\caption{Time series to illustrate the rules applied when labeling experimental data manually (expert criteria). (A) corresponds to \textit{Rule~a}, (B) to \textit{Rule b}, (C) to \textit{Rule c} and (D) to \textit{Rule d} (see the text for more details). In green we have the points that correspond to the time series, and we have joined such points with black segments for the ease of viewing.}
		\label{ExampleSamples}
	\end{center}
\end{figure}

To check if the analysis of chaotic dynamics has been successful, we need to label the experimental dataset manually according to expert criteria. In general, we consider the asymptotic behavior to label the data. In particular, the rules taken into account for such classification are as follows:

\begin{description}
	\item[{{Rule a.}}] If a time series seems to have long chaotic transient behavior (more than a quarter of the length) and only some regularity can be inferred at the end (see Figure~\ref{ExampleSamples}(A) for an example), it is considered that the behavior is chaotic. In this case, we cannot ensure that the asymptotic behavior is regular as not enough information is provided.
	
	\item[{{Rule b.}}] If a time series seems to have short chaotic transient behavior (less than a quarter of the length) and some regularity later (as in Figure~\ref{ExampleSamples}(B)), it is classified by the expert as regular.
	
	\item[{{Rule c.}}] If some regular windows are present in a general chaotic behavior (see Figure~\ref{ExampleSamples}(C)), chaotic behavior is assigned.
	
	\item[{{Rule d.}}] As we work with experimental data, the dynamics corresponding to equilibrium points are not constant values over time. We consider that a time series is an equilibrium point or represents regular behavior if when transient has ended (that is, in the last three quarters according to \textit{Rule b}), the range of values taken by the variable in the original range before normalization is small (a difference less than $25$ approximately) or a monotonically increasing or decreasing behavior can be inferred (see Figure~\ref{ExampleSamples}(D) for an example).
\end{description}
Taking into account these rules, in Dataset 1, $13$ of the $52$ samples have chaotic behavior, and the remaining $39$ are regular. And in Dataset 2, we have that $20$ of the $24$ samples are regular and just $4$ are chaotic.

\begin{figure*}[h!]
	\begin{center}
		\includegraphics[width=0.75\textwidth]{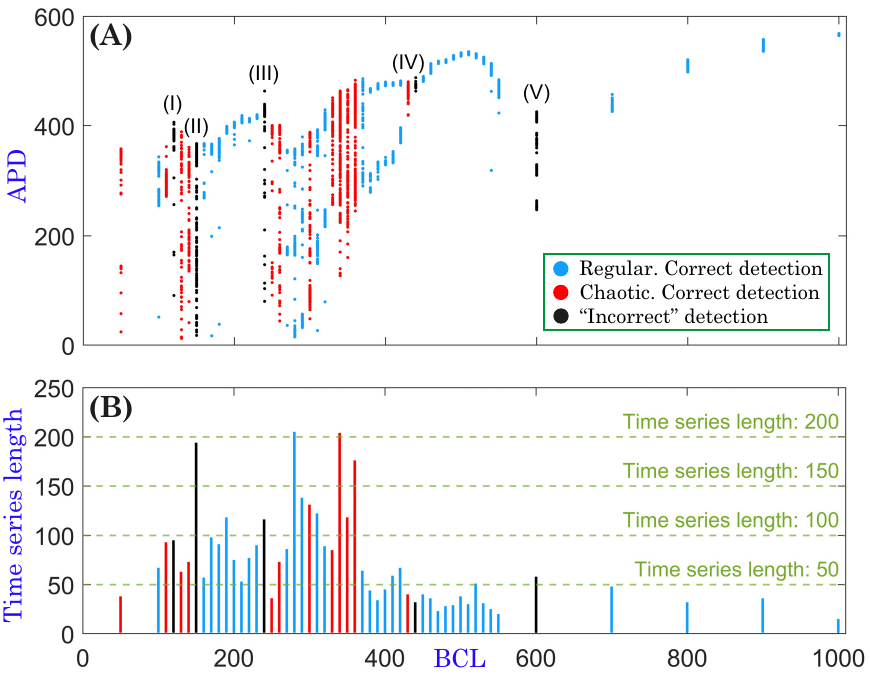}
		\caption{Chaos analysis of experimental data of frog heart dynamics from Dataset 1 using the DL algorithm of Section~\ref{sec:2}. (A) Results of the chaos analysis of experimental time series. (B) Time series length (that is, number of time points). In both panels, colors correspond to the DL classification: Blue is used for correct regular detection, red for correct chaotic detection, and black for ``incorrect'' detection. (I)-(V) correspond to the ``incorrect'' detections and the time series are drawn in Figure~\ref{Fails}.}
		\label{ExpResults}
	\end{center}
\end{figure*}

First of all, we use the proposed algorithm to analyze chaotic dynamics in Dataset 1. In Figure~\ref{ExpResults}(A), the DL chaos analysis of experimental data of heart dynamics is depicted. In blue we have the samples that have been correctly detected as regular by the algorithm (compared with the expert criteria), in red those correctly classified as chaotic, and in black the ``incorrectly'' detected ones (that is, false regular and false chaotic detections). As we can see, just $5$ of the $52$ samples have been incorrectly classified ($9.615\%$ of the samples). That is, the DL algorithm for chaos analysis has been successful in $90.385\%$ of the dataset ($47$ of $52$ samples). As our dataset is not balanced (we have more regular than chaotic samples), it is important to compute the accuracy chaotic and accuracy regular. In particular, such values are $92.308\%$ ($12$ of $13$ samples) and $89.744\%$ ($35$ of $39$ time series), respectively. Notice that all the accuracy values are around $90\%$ of success, and it seems that the DL algorithm for chaos analysis of heart time series has worked properly. In Table~\ref{Table_ExpResults} we have summarized such accuracy results.

\begin{table}[h!]
	\resizebox{.48\textwidth}{!}{
	\begin{tabular}{|c||c|}
		\hline
		DATASET 1 & \cellcolor{gray!60}{\rm{\textbf{Experimental Results}}}\\
		\hline
		\hline
		\cellcolor{gray!20}\centering {\rm{\textbf{Accuracy (\%)}}} & $90.385\,\%$ {\small{($47$ of $52$ samples)}}\\
		\hline
		\cellcolor{gray!20}\centering {\rm{\textbf{Accuracy Chaotic (\%)}}} & $92.308\,\%$ {\small{($12$ of $13$ samples)}}\\
		\hline
		\cellcolor{gray!20}\centering {\rm{\textbf{Accuracy Regular (\%)}}} & $89.744\,\%$ {\small{($35$ of $39$ samples)}}\\
		\hline
	\end{tabular}}
	\caption{Results of the DL chaos analysis of the experimental data (frog heart dynamics) of Dataset 1 with the network obtained in the algorithm.}
	\label{Table_ExpResults}
\end{table}

In Figure~\ref{ExpResults}(B) we have represented the length, number of data points, of each experimental time series ($x$-axis correspond to the BCL value of the sample, and the height of the line is the length). Colors correspond to the network classification: blue for correct regular detection, red for correct chaotic detection, and black for ``incorrect detection''. The green dashed horizontal lines indicate length values $50$, $100$, $150$ and $200$, and have been added to facilitate visualization. As already mentioned when describing the experimental dataset, the time series have different lengths (minimum length $15$ and maximum length $205$). In particular, $40.385\%$ of the samples ($21$ of $52$) contain at most $50$ time points, and $80.769\%$ ($42$ of $52$ samples) have at most $100$. Notice that most of the samples are quite short, and this can complicate the task as not enough information can be provided to the network. However, it seems that the applied DL algorithm for chaos analysis is able to detect properly most of the regular and chaotic behavior regardless of the length.

\begin{figure}
	\includegraphics[width=0.5\textwidth]{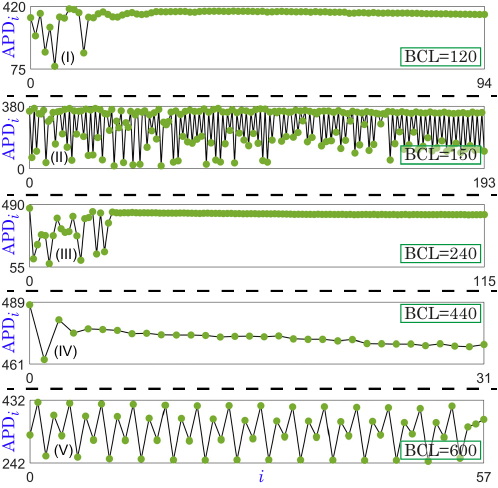}
	\caption{Time series that have been ``incorrectly'' detected in Dataset 1. They correspond to samples (I)-(V) in black in Figure~\ref{ExpResults}. In green we have the points that correspond to the time series and we have joined such points with black segments for a better visualization of the behavior.}
	\label{Fails}
\end{figure}

Let us analyze in detail the ``incorrectly'' detected samples (in black in Figure~\ref{ExpResults}). We have represented them in Figure~\ref{Fails}. In green we have the points that correspond to the time series, and we have connected such points with black segments for a better visualization of the behavior. Notice that not all the samples have the same length, that is, the scale of $x$-axis is different for each time series. As can be seen in Figure~\ref{Fails}, samples (I), (III) and (IV) have a similar behavior: Short chaotic transient dynamics, asymptotically converging to an equilibrium point (regular). According to \textit{Rule b} that we have used for expert classification of the experimental time series, the samples have been labeled as regular. The ANN detected them as chaotic. The whole time series are chaotic (as DL detected), but dynamically their asymptotic behavior can be considered as regular (as we labeled them). Sample (II), as can be seen in Figure~\ref{Fails}, seems to have some periodicity at the end, with long chaotic transient. Therefore, applying \textit{Rule a} of the expert classification, the time series has been labeled as chaotic. However, the algorithm detected it as regular. The fifth incorrect detection corresponds to sample (V). As seen in Figure~\ref{Fails}, it can be difficult to classify such an orbit. In fact, this data seems to correspond to an orbit inside or just after a period-doubling cascade,  so it can be difficult to classify it as a periodic orbit with a high period, a  quasi-periodic orbit or a Feigenbaum chaotic orbit. As it seems to follow some periodicity along all the time points, it has been labeled as regular. The network detected it as chaotic with a score (``probability'') of $0.625$ for this class (that is, with a high uncertainty). Note that all these ``incorrectly'' classified time series have regular and chaotic patterns, and so any classification cannot be accurate as there are too few data points to correctly classify them. Therefore, the results of the ANN cannot really be considered as a false result (and for this reason we have used quotation marks in the word incorrect when we refer to them).

\begin{figure}[h!]
	\begin{center}
		\includegraphics[width=0.5\textwidth]{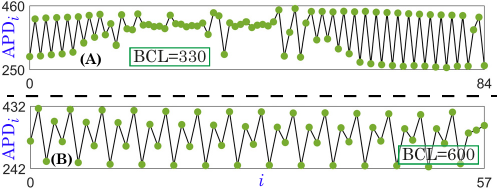}
		\caption{Time series of Dataset 1 with detection score (``probability'') less than $2/3$ for the winning class. In green we have the dots that correspond to the time series and we have connected them with black segments for the ease of viewing.}
		\label{InDoubtSamples}
	\end{center}
\end{figure}

\begin{Remark}
	Notice that in the case of sample (V) (see Figures~\ref{ExpResults} and \ref{Fails}), the network assigned a score $0.625$ for chaotic class and a score $0.375$ for regular class. The values are closer to $0.5$ than to $1$, what means that the ANN is not sure about its decision. If some human supervision wants to be added to the algorithm defined in Section~\ref{sec:2}, we can add a control to \textit{Step 4}. It would be defined as follows: If the higher score of one sample for both categories is lower than $2/3$, such data point should be revised by an expert as the detection score is close to the classification border ($0.5$ is the threshold to distinguish between regular and chaotic). Adding this control in our experimental dataset, just $3.486\%$ of the samples ($2$ of $52$ samples) should be revised. These two samples have been represented in Figure~\ref{InDoubtSamples}. As already indicated, one of these samples (sample (B)) corresponds to one of the samples ``incorrectly'' detected (see sample (V) of Figure~\ref{Fails}). The other one (sample (A)) was correctly classified by the algorithm. The network assigned a score $0.592$ for the chaotic class and $0.408$ for regular class. Notice that the time series seems to follow some periodicity at the beginning and the end of the recordings but in between it presents a chaotic behavior, so the doubt of the network could be totally justified. After this expert supervision, the final accuracy for the whole dataset would be $92.308\%$ ($48$ of the $52$ samples), with accuracy regular increasing also to this value ($36$ of $39$ samples).
\end{Remark}

\begin{table}[h!]
	\resizebox{.48\textwidth}{!}{
		\begin{tabular}{|c||c|}
			\hline
			DATASET 2 & \cellcolor{gray!60}{\rm{\textbf{Experimental Results}}}\\
			\hline
			\hline
			\cellcolor{gray!20}\centering {\rm{\textbf{Accuracy (\%)}}} & $91.667\,\%$ {\small{($22$ of $24$ samples)}}\\
			\hline
			\cellcolor{gray!20}\centering {\rm{\textbf{Accuracy Chaotic (\%)}}} & $75.000\,\%$ {\small{($3$ of $4$ samples)}}\\
			\hline
			\cellcolor{gray!20}\centering {\rm{\textbf{Accuracy Regular (\%)}}} & $95.000\,\%$ {\small{($19$ of $20$ samples)}}\\
			\hline
	\end{tabular}}
	\caption{Results of the DL chaos analysis of the experimental data (frog heart dynamics) of Dataset 2 with the network obtained in the algorithm.}
	\label{Table_ExpResults2}
\end{table}

\begin{figure}
	\begin{center}
		\includegraphics[width=0.48\textwidth]{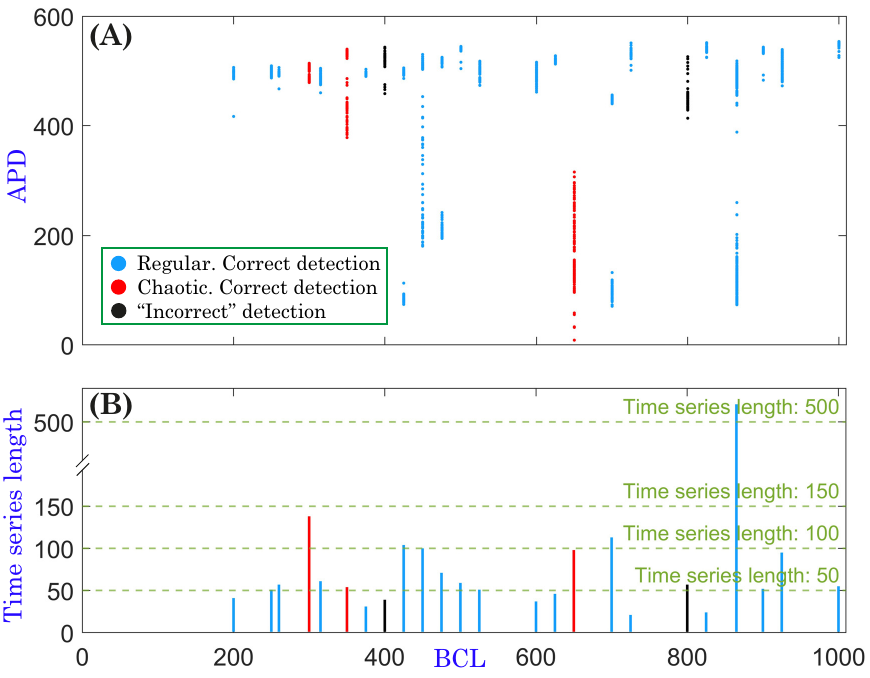}
		\caption{Chaos analysis of experimental data of frog heart dynamics from Dataset 2 using the DL algorithm of Section~\ref{sec:2}. (A) Results of the chaos analysis of experimental time series. (B) Time series length (that is, number of time points). In both panels, colors correspond to the DL classification: Blue is used for correct regular detection, red for correct chaotic detection, and black for ``incorrect'' detection.}
		\label{ExpResults2}
	\end{center}
\end{figure}

Finally, we apply the proposed algorithm to Dataset 2. In Figure~\ref{ExpResults2}(A), the DL analysis of chaotic dynamics of the experimental heart data is represented following the same code of colors of the study of Dataset 1 (blue for regular samples correctly detected comparing with expert criteria, red for correctly classified chaotic time series, and black for ``incorrectly'' classified signals). As summarized in Table~\ref{Table_ExpResults2}, $91.667\%$ of the samples have been classified properly ($22$ of $24$ time series), in fact, we obtain a value of $95\%$ for accuracy regular ($19$ of $20$ samples) and $75\%$ for accuracy chaotic ($3$ of $4$ samples). Notice that this accuracy chaotic value is not really high, but we have to take into account that there are only $4$ chaotic samples in the dataset and just $1$ has been misclassified (that is, small set of samples in which just one error strongly degrades the results). Only two samples have been ``incorrectly'' classified in the whole dataset compared to the expert detection, both with a short number of points (length) as can be seen in Figure~\ref{ExpResults2}(B). This panel illustrates that except for one long sample, the remaining signals have in general less than $100$ points.

\subsection{Brief Comparison with Other Standard Techniques}
\label{subsec:33}

In this subsection we perform a brief study of the results provided by different standard techniques in the analysis of chaotic dynamics in the experimental frog heart signals of Dataset 1. This will allow us to compare with the performance of the algorithm proposed in the present work. The standard techniques that we apply are the Permutation Entropy (PE)~\cite{BandtPompe02,Riedl13} (Python package antropy~\cite{vallat2025antropy} has been used for the implementation) and the 0\,-1 test for chaos~\cite{gottwald20160} (MATLAB implementation in~\cite{matthews2025chaos} has been applied).

Permutation Entropy is considered a noisy but robust technique that measures the complexity of data even in the presence of dynamical and observational noise. The {0\,-1} test for chaos is a powerful and widely applied tool to distinguish between chaotic and regular behavior in dynamical systems. As shown in Figures~\ref{LM_image} and~\ref{BR_image}, both techniques provide accurate and meaningful results when applied to mathematical models, and they have been widely used in the literature~\cite{EYEBEFOUDA2016259, halfar2020dynamical}. Therefore, we consider them suitable tools to compare with our algorithm.

\begin{figure}[h!]
	\begin{center}
		\includegraphics[width = 0.48\textwidth]{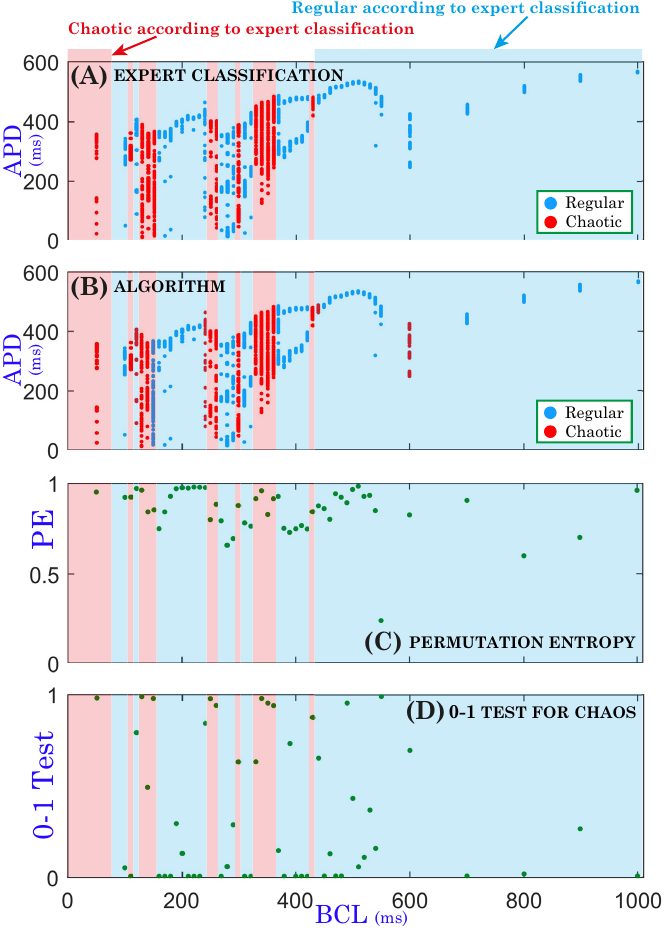}
		\caption{Analysis of chaotic dynamics for the experimental data (frog heart signals) of Dataset 1 using different techniques. (A) Bifurcation diagram with the results provided by the expert classification. Blue corresponds to regular and red to chaotic dynamics. (B) Bifurcation diagram with the results provided by the DL algorithm proposed in this paper (it corresponds to panel (A) of Figure~\ref{ExpResults}. Blue corresponds to regular and red to chaotic dynamics. (C) Permutation entropy (PE). (D) 0\,-1 test for chaos. Color shading follows the behavior obtained with the expert classification that we consider as ground truth to compare with the remaining techniques.}
		\label{Exp_Other_Techniques}
	\end{center}
\end{figure}

In Figure~\ref{Exp_Other_Techniques} we have the analysis of chaotic dynamics provided by the expert classification in panel (A) (it is based on \textit{Rules a-d} explained in Subsection~\ref{subsec:32} and it is considered as the ground truth), by our algorithm in panel (B) (this panel coincides with panel (A) of Figure~\ref{ExpResults}), by the Permutation Entropy (PE) technique in panel (C), and by the 0\,-1 test for chaos in panel (D). Blue and red shading across all the panels correspond to the results provided by the expert classification that we consider as ground truth (blue corresponds to regular label and red to chaotic one). The Permutation Entropy technique depends on two parameters, the permutation order $n$ and the delay $\tau$, we have set them to $3$ and $1$, respectively, following the recommendations in~\cite{Riedl13}. For the permutation order, it is suggested to use the maximum $n$ value such that the length of the time series is larger than $5n!$. As we have a large variability in the time series length across our experimental dataset, we consider $n=3$ that satisfies the criterion for most of the samples. For the delay $\tau$, we consider the recommendation indicated for one-dimensional dynamical systems, as working with single-variable time series and the corresponding mathematical model (APD heart map) is one-dimensional.

If we focus on panel (C) of Figure~\ref{Exp_Other_Techniques} corresponding to the (normalized) Permutation Entropy, at first sight we can see that values greater than 0.5 have been obtained for almost all the samples. This seems to provide not meaningful results (or at least results that are not easy to interpret). If we study the plot carefully, we can see that, for instance, for the regular samples in the fourth and fifth blue shadings (from left to right), the complexity is a bit lower than for the samples in the neighboring red shading regions, showing that they are more regular and slightly aligned with the expert classification. However, this difference does not seem to be remarkable. On the other hand, we can see that most of the Permutation Entropy values for regular samples in the third and last blue shading regions are larger than those for the chaotic samples in the neighboring regions, providing results that would not allow us to classify the dynamical behavior easily. Therefore, we can infer that Permutation Entropy does not provide proper or easy to interpret results for our particular dataset.

If we take a look on panel (D) of Figure~\ref{Exp_Other_Techniques} corresponding to the 0\,-1 test for chaos, we can see that for several samples (around $20$ time series, that is, more than one third of the dataset) we obtain intermediate values, neither close to $0$ nor to $1$, not providing a dynamical classification for them. For the remaining ones, except for a couple of samples in the most-right blue shading, the classification is adequate. In this case, we can see that the 0\,-1 test for chaos cannot provide proper results in a wide part of the dataset because of the nature of the samples (short time series).

Comparing the performance of the standard techniques (Permutation Entropy in panel (C) of Figure~\ref{Exp_Other_Techniques} and 0\,-1 test for chaos in panel (D)) and our algorithm (see panel (B) of Figure~\ref{Exp_Other_Techniques} and Subsection~\ref{subsec:32} for more details), we can see that our proposed tool to analyze chaotic dynamics in experimental data is the more accurate (at least in our dataset). This could be mainly due to the nature of the data (very short noisy time series). That is, as it is also observed in literature, the classical techniques usually need sufficiently enough long time series, but this is not always possible like occurs in heart data. Therefore, new techniques like the one introduced here can help in the analysis of this kind of experimental data.

\section{Conclusions}
\label{sec:Conclusions}
There is an increasing amount of experimental data in practice, so automatic techniques that would allow to classify them properly are necessary. Real-world data often has some drawbacks, such as noisy and short recordings, or limited amount of samples. In addition, not all the variables of the real-world system can be available for recording. In this paper, we propose a Deep Learning-based algorithm to tackle the chaos analysis of biological time series. The algorithm is divided into different steps. We consider an Artificial Neural Network architecture based on Long Short-Term Memory cells with a final classification layer. The training data is taken from a basic and generic discrete dynamical system: the Logistic Map. Once $10$ Artificial Neural Networks with that same architecture but different initialization of the trainable parameters have been trained with such generic data, a mathematical model with the same nature as the experimental data is used to perform a test (in our case of study, the Beeler-Reuter APD heart map). From this analysis, applying some selection criteria, one of the trained Artificial Neural Networks is selected. Such network seems to be adequate to perform the analysis of chaotic dynamics of biological time series (in our case, of frog heart signals).

The algorithm is tested on a set of experimental data obtained from frog heart signals in the laboratory of Professor Flavio H. Fenton. The accurate results of the paper highlights that the combination of Deep Learning techniques and mathematical models can be useful to face the analysis of chaotic dynamics in experimental data even when it consists of short time series (where other techniques, such as the computation of the maximum Lyapunov exponent would not work properly). In fact, the algorithm shows that Artificial Neural Networks trained in generic data can be used for particular problems once the most appropriate one has been selected using data from a problem similar or of the same nature as the particular problem we want to address. Moreover, this is a \textit{de facto} proof that most dynamical phenomena are quite general and universal. However, the proposed algorithm has some limitations. One of them stems from the use of Deep Learning. Currently, this technique lacks general interpretability, which introduces a disadvantage into our algorithm that is not typically present in classical algorithms. On the other hand, one of the crucial points of the algorithm is the use of a mathematical model that shares the nature of the experimental data. Although rare, it is possible that no suitable model is available for the type of experimental data to be analyzed. As future work, the authors plan to investigate how emerging branches of Deep Learning that incorporate physical information, such as Physics-Informed Neural Networks (PINNs) which are expected to provide more interpretability, could be applied to the analysis of experimental data.

\section*{Acknowledgments}

RB and CMC have been supported by the Spanish Research projects PID2021-122961NB-I00 and PID2024-156032NB-I00, and the European Regional Development Fund and Diputaci\'on General de Arag\'on E24-23R. RB has been supported by the European Regional Development Fund and Diputaci\'on General de Arag\'on LMP94-21. CMC has been supported by Spanish Research project PID2022-140556OB-I00 and by Ministerio de Universidades de España with an FPU grant (FPU20/04039). FHF, MH, CH and MJT have been supported by NIH, United States grant 2R01HL143450.


\section*{Declaration of Competing Interest}

There is no conflict of interest between the authors and other persons organizations.

\section*{Data Availability}

Data available on request from the authors.

\bibliographystyle{elsarticle-num}

\begin{thebibliography}{10}
	\expandafter\ifx\csname url\endcsname\relax
	\def\url#1{\texttt{#1}}\fi
	\expandafter\ifx\csname urlprefix\endcsname\relax\def\urlprefix{URL }\fi
	\expandafter\ifx\csname href\endcsname\relax
	\def\href#1#2{#2} \def\path#1{#1}\fi
	
	\setlength{\parskip}{-0.5pt}
	
	{\small{\bibitem{hastings2000alternans}
	H.~M. Hastings, F.~H. Fenton, S.~J. Evans, O.~Hotomaroglu, J.~Geetha,
	K.~Gittelson, J.~Nilson, A.~Garfinkel, Alternans and the onset of ventricular
	fibrillation, Physical Review E 62~(3) (2000) 4043.
	
	\bibitem{xie2007dispersion}
	Y.~Xie, G.~Hu, D.~Sato, J.~N. Weiss, A.~Garfinkel, Z.~Qu, Dispersion of
	refractoriness and induction of reentry due to chaos synchronization in a
	model of cardiac tissue, Physical Review Letters 99~(11) (2007) 118101.
	
	\bibitem{munoz2021controllability}
	L.~M. Mu{\~n}oz, M.~O. Ampofo, E.~M. Cherry, Controllability of voltage-and
	calcium-driven cardiac alternans in a map model, Chaos: An Interdisciplinary
	Journal of Nonlinear Science 31~(2) (2021).
	
	\bibitem{stenzinger2023cardiac}
	R.~Stenzinger, T.~Scalvin, P.~Morelo, M.~Tragtenberg, Cardiac behaviors and
	chaotic arrhythmias in the {H}indmarsh--{R}ose model, Chaos, Solitons \&
	Fractals 175 (2023) 113983.
	
	\bibitem{wang2024intracellular}
	X.~Wang, J.~Landaw, Z.~Qu, Intracellular ion accumulation in the genesis of
	complex action potential dynamics under cardiac diseases, Physical Review E
	109~(2) (2024) 024410.
	
	\bibitem{Garfinkel1992}
	A.~Garfinkel, M.~Spano, W.~Ditto, J.~Weiss, Controlling cardiac chaos, Science
	(New York, N.Y.) 257 (1992) 1230--5.
	
	\bibitem{skinner1990chaos}
	J.~E. Skinner, A.~L. Goldberger, G.~Mayer-Kress, R.~E. Ideker, Chaos in the
	heart: {I}mplications for clinical cardiology, Bio/technology 8~(11) (1990)
	1018--1024.
	
	\bibitem{710541}
	N.~Thakor, Chaos in the heart: {S}ignals and models, in: Proceedings of the
	1998 2nd International Conference Biomedical Engineering Days, 1998, pp.
	11--18.
	
	\bibitem{goldberger1991normal}
	A.~L. Goldberger, Is the normal heartbeat chaotic or homeostatic?, Physiology
	6~(2) (1991) 87--91.
	
	\bibitem{poon1997decrease}
	C.-S. Poon, C.~K. Merrill, Decrease of cardiac chaos in congestive heart
	failure, Nature 389~(6650) (1997) 492--495.
	
	\bibitem{goldberger1986some}
	A.~L. Goldberger, V.~Bhargava, B.~J. West, A.~J. Mandell, Some observations on
	the question: {I}s ventricular fibrillation “chaos”?, Physica D:
	Nonlinear Phenomena 19~(2) (1986) 282--289.
	
	\bibitem{weiss1999chaos}
	J.~N. Weiss, A.~Garfinkel, H.~S. Karagueuzian, Z.~Qu, P.-S. Chen, Chaos and the
	transition to ventricular fibrillation: a new approach to antiarrhythmic drug
	evaluation, Circulation 99~(21) (1999) 2819--2826.
	
	\bibitem{gupta2021chaos}
	V.~Gupta, M.~Mittal, V.~Mittal, Chaos theory and {ARTFA}: emerging tools for
	interpreting {ECG} signals to diagnose cardiac arrhythmias, Wireless Personal
	Communications 118~(4) (2021) 3615--3646.
	
	\bibitem{chialvo1987non}
	D.~R. Chialvo, J.~Jalife, Non-linear dynamics of cardiac excitation and impulse
	propagation, Nature 330~(6150) (1987) 749--752.
	
	\bibitem{chialvo1990low}
	D.~R. Chialvo, R.~F. Gilmour~Jr, J.~Jalife, Low dimensional chaos in cardiac
	tissue, Nature 343~(6259) (1990) 653--657.
	
	\bibitem{gizzi2013effects}
	A.~Gizzi, E.~M. Cherry, R.~F. Gilmour~Jr, S.~Luther, S.~Filippi, F.~H. Fenton,
	Effects of pacing site and stimulation history on alternans dynamics and the
	development of complex spatiotemporal patterns in cardiac tissue, Frontiers
	in Physiology 4 (2013) 71.
	
	\bibitem{iravanian2023complex}
	S.~Iravanian, I.~Uzelac, A.~D. Shah, M.~J. Toye, M.~S. Lloyd, M.~A. Burke,
	M.~A. Daneshmand, T.~S. Attia, J.~D. Vega, M.~F. El-Chami, F.~M. Merchant,
	E.~M. Cherry, N.~K. Bhatia, F.~H. Fenton, Complex repolarization dynamics in
	ex vivo human ventricles are independent of the restitution properties,
	Europace 25~(12) (2023) euad350.
	
	\bibitem{lilienkamp2022taming}
	T.~Lilienkamp, U.~Parlitz, S.~Luther, Taming cardiac arrhythmias: Terminating
	spiral wave chaos by adaptive deceleration pacing, Chaos: An
	Interdisciplinary Journal of Nonlinear Science 32~(12) (2022).
	
	\bibitem{detal2022terminating}
	N.~DeTal, A.~Kaboudian, F.~H. Fenton, Terminating spiral waves with a single
	designed stimulus: Teleportation as the mechanism for defibrillation,
	Proceedings of the National Academy of Sciences 119~(24) (2022) e2117568119.
	
	\bibitem{tyler2024experimental}
	S.~A. Tyler, D.~Mersing, F.~H. Fenton, M.~R. Tinsley, K.~Showalter,
	Experimental studies of spiral wave teleportation in a light sensitive
	{B}elousov--{Z}habotinsky system, Chaos: An Interdisciplinary Journal of
	Nonlinear Science 34~(9) (2024).
	
	\bibitem{ji2017synchronization}
	Y.~C. Ji, I.~Uzelac, N.~Otani, S.~Luther, R.~F. Gilmour~Jr, E.~M. Cherry, F.~H.
	Fenton, Synchronization as a mechanism for low-energy anti-fibrillation
	pacing, Heart rhythm 14~(8) (2017) 1254--1262.
	
	\bibitem{argyris1994exploration}
	J.~H. Argyris, G.~Faust, M.~Haase, An Exploration of Chaos: {A}n Introduction
	for Natural Scientists and Engineers, North Holland, 1994.
	
	\bibitem{wolf1985determining}
	A.~Wolf, J.~B. Swift, H.~L. Swinney, J.~A. Vastano, Determining {L}yapunov
	exponents from a time series, Physica D: Nonlinear Phenomena 16~(3) (1985)
	285--317.
	
	\bibitem{rosenstein1993practical}
	M.~T. Rosenstein, J.~J. Collins, C.~J. De~Luca, A practical method for
	calculating largest {L}yapunov exponents from small data sets, Physica D:
	Nonlinear Phenomena 65~(1-2) (1993) 117--134.
	
	\bibitem{BandtPompe02}
	C.~Bandt, B.~Pompe,
	\href{https://link.aps.org/doi/10.1103/PhysRevLett.88.174102}{Permutation
		entropy: A natural complexity measure for time series}, Phys. Rev. Lett. 88
	(2002) 174102.
	\newblock \href {https://doi.org/10.1103/PhysRevLett.88.174102}
	{\path{doi:10.1103/PhysRevLett.88.174102}}.
	\newline\urlprefix\url{https://link.aps.org/doi/10.1103/PhysRevLett.88.174102}
	
	\bibitem{Amigo10}
	J.~Amig\'o, Permutation Complexity in Dynamical Systems, Springer Berlin
	Heidelberg, 2010.
	
	\bibitem{stosic2022generalized}
	D.~Stosic, D.~Stosic, T.~Stosic, B.~Stosic, Generalized weighted permutation
	entropy, Chaos: An Interdisciplinary Journal of Nonlinear Science 32~(10)
	(2022).
	
	\bibitem{Riedl13}
	M.~Riedl, A.~M\"{u}ller, N.~Wessel, Practical considerations of permutation
	entropy, The European Physical Journal Special Topics 222 (2013) 249--262.
	\newblock \href {https://doi.org/10.1140/epjst/e2013-01862-7}
	{\path{doi:10.1140/epjst/e2013-01862-7}}.
	
	\bibitem{gottwald20160}
	G.~A. Gottwald, I.~Melbourne, The 0\,-1 Test for Chaos: A Review, Springer
	Berlin Heidelberg, Berlin, Heidelberg, 2016, pp. 221--247.
	\newblock \href {https://doi.org/10.1007/978-3-662-48410-4_7}
	{\path{doi:10.1007/978-3-662-48410-4_7}}.
	
	\bibitem{barrio2007three}
	R.~Barrio, S.~Serrano, A three-parametric study of the {L}orenz model, Physica
	D: Nonlinear Phenomena 229~(1) (2007) 43--51.
	
	\bibitem{EYEBEFOUDA2016259}
	J.~A. {Eyebe Fouda}, B.~Bodo, G.~M. Djeufa, S.~L. Sabat, Experimental chaos
	detection in the {D}uffing oscillator, Communications in Nonlinear Science
	and Numerical Simulation 33 (2016) 259--269.
	\newblock \href {https://doi.org/https://doi.org/10.1016/j.cnsns.2015.09.011}
	{\path{doi:https://doi.org/10.1016/j.cnsns.2015.09.011}}.
	
	\bibitem{barrio2021dynamical}
	R.~Barrio, M.~Mart{\'\i}nez, E.~Pueyo, S.~Serrano, Dynamical analysis of
	{E}arly {A}fterdepolarization patterns in a biophysically detailed cardiac
	model, Chaos: An Interdisciplinary Journal of Nonlinear Science 31~(7)
	(2021).
	
	\bibitem{halfar2020dynamical}
	R.~Halfar, Dynamical properties of {B}eeler-{R}euter cardiac cell model with
	respect to stimulation parameters, International Journal of Computer
	Mathematics 97~(1-2) (2020) 498--507.
	
	\bibitem{shintani2024observation}
	S.~A. Shintani, Observation of sarcomere chaos induced by changes in calcium
	concentration in cardiomyocytes, Biophysics and Physicobiology 21~(1) (2024)
	e210006.
	
	\bibitem{islam2025dynamic}
	M.~A. Islam, I.~R. Hassan, P.~Ahmed, Dynamic complexity of fifth-dimensional
	{H}enon map with {L}yapunov exponent, permutation entropy, bifurcation
	patterns and chaos, Journal of Computational and Applied Mathematics 466
	(2025) 116547.
	
	\bibitem{ma2025using}
	X.~Ma, G.~Litak, S.~Zhou, Using 0--1 test to diagnose periodic and chaotic
	motions of nonlinear vortex-induced vibration energy harvesters, Chaos,
	Solitons \& Fractals 192 (2025) 116036.
	
	\bibitem{Toker20}
	D.~Toker, F.~Sommer, M.~A. D’Esposito, A simple method for detecting chaos in
	nature, Communications Biology 3~(article number 11) (2020).
	
	\bibitem{palus1998chaotic}
	M.~Palu{\v{s}}, Chaotic measures and real-world systems: Does the {L}yapunov
	exponent always measure chaos?, in: H.~Kantz, J.~Kurths, G.~Mayer-Kress
	(Eds.), Nonlinear Analysis of Physiological Data, Springer Berlin Heidelberg,
	Berlin, Heidelberg, 1998, pp. 49--66.
	
	\bibitem{zanin2021ordinal}
	M.~Zanin, F.~Olivares, Ordinal patterns-based methodologies for distinguishing
	chaos from noise in discrete time series, Communications Physics 4~(1) (2021)
	190.
	
	\bibitem{kumar09}
	A.~Kumar, Light propagation through biological tissue: comparison between
	{M}onte {C}arlo simulation and deterministic models, International Journal of
	Biomedical Engineering and Technology 2~(4) (2009) 344--351.
	\newblock \href {https://doi.org/10.1504/IJBET.2009.027798}
	{\path{doi:10.1504/IJBET.2009.027798}}.
	
	\bibitem{kumar25}
	A.~Kumar, A.~Pal Singh~Chauhan, Robust feature extraction from omnidirectional
	outdoor images for computer vision applications, International Journal of
	Instrumentation and Measurement 10 (2025) 8--13.
	\newblock \href {https://doi.org/10.1504/IJBET.2009.027798}
	{\path{doi:10.1504/IJBET.2009.027798}}.
	
	\bibitem{CGRV22}
	A.~Celletti, C.~Gales, V.~Rodriguez-Fernandez, M.~Vasile, Classification of
	regular and chaotic motions in {H}amiltonian systems with {D}eep {L}earning,
	Scientific Reports 12~(1) (2022) 1--12.
	
	\bibitem{ChaosDetCody23}
	R.~Barrio, {\'A}.~Lozano, A.~Mayora-Cebollero, C.~Mayora-Cebollero, A.~Miguel,
	A.~Ortega, S.~Serrano, R.~Vigara, {{D}eep {L}earning for chaos detection},
	Chaos: An Interdisciplinary Journal of Nonlinear Science 33~(7) (2023)
	073146.
	
	\bibitem{pathak2017using}
	J.~Pathak, Z.~Lu, B.~R. Hunt, M.~Girvan, E.~Ott, Using {M}achine {L}earning to
	replicate chaotic attractors and calculate {L}yapunov exponents from data,
	Chaos: An Interdisciplinary Journal of Nonlinear Science 27~(12) (2017).
	
	\bibitem{mayora2024full}
	C.~Mayora-Cebollero, A.~Mayora-Cebollero, Álvaro Lozano, R.~Ba-rrio, Full
	{L}yapunov exponents spectrum with {D}eep {L}earning from single-variable
	time series, Physica D: Nonlinear Phenomena 472 (2025) 134510.
	
	\bibitem{shahi2022machine}
	S.~Shahi, F.~H. Fenton, E.~M. Cherry, A {M}achine-{L}earning approach for
	long-term prediction of experimental cardiac action potential time series
	using an {A}utoencoder and {E}cho {S}tate {N}etworks, Chaos: An
	Interdisciplinary Journal of Nonlinear Science 32~(6) (2022).
	
	\bibitem{shahi2022prediction}
	S.~Shahi, F.~H. Fenton, E.~M. Cherry, Prediction of chaotic time series using
	{R}ecurrent {N}eural {N}etworks and {R}eservoir {C}omputing techniques: A
	comparative study, Machine Learning with Applications 8 (2022) 100300.
	
	\bibitem{nolasco1968graphic}
	J.~Nolasco, R.~W. Dahlen, A graphic method for the study of alternation in
	cardiac action potentials, Journal of Applied Physiology 25~(2) (1968)
	191--196.
	
	\bibitem{guevara1984electrical}
	M.~Guevara, G.~Ward, A.~Shrier, L.~Glass, Electrical alternans and period
	doubling bifurcations, Computing in Cardiology 562 (1984) 167--170.
	
	\bibitem{franz1988cycle}
	M.~R. Franz, C.~D. Swerdlow, L.~B. Liem, J.~Schaefer, et~al., Cycle length
	dependence of human action potential duration in vivo. effects of single
	extrastimuli, sudden sustained rate acceleration and deceleration, and
	different steady-state frequencies., The Journal of Clinical Investigation
	82~(3) (1988) 972--979.
	
	\bibitem{szigligeti1998intracellular}
	P.~Szigligeti, T.~B\'any\'asz, J.~Magyar, G.~Szigeti, Z.~Papp, A.~Varr\'o,
	P.~N\'an\'asi, Intracellular calcium and electrical restitution in mammalian
	cardiac cells, Acta Physiol Scand. 163~(2) (1998) 139--147.
	
	\bibitem{watanabe1995biphasic}
	M.~Watanabe, N.~F. Otani, R.~F. Gilmour~Jr, Biphasic restitution of action
	potential duration and complex dynamics in ventricular myocardium,
	Circulation Research 76~(5) (1995) 915--921.
	
	\bibitem{qu1997spatiotemporal}
	Z.~Qu, J.~N. Weiss, A.~Garfinkel, Spatiotemporal chaos in a simulated ring of
	cardiac cells, Physical Review Letters 78~(7) (1997) 1387.
	
	\bibitem{fenton2002multiple}
	F.~H. Fenton, E.~M. Cherry, H.~M. Hastings, S.~J. Evans, Multiple mechanisms of
	spiral wave breakup in a model of cardiac electrical activity, Chaos: An
	Interdisciplinary Journal of Nonlinear Science 12~(3) (2002) 852--892.
	
	\bibitem{fenton1999memory}
	F.~H. Fenton, S.~J. Evans, H.~M. Hastings, Memory in an excitable medium: {A}
	mechanism for spiral wave breakup in the low-excitability limit, Physical
	Review Letters 83~(19) (1999) 3964.
	
	\bibitem{tolkacheva2003condition}
	E.~G. Tolkacheva, D.~G. Schaeffer, D.~J. Gauthier, W.~Krassowska, Condition for
	alternans and stability of the 1:1 response pattern in a “memory” model
	of paced cardiac dynamics, Physical Review E 67~(3) (2003) 031904.
	
	\bibitem{fox2002period}
	J.~J. Fox, E.~Bodenschatz, R.~F. Gilmour~Jr, Period-doubling instability and
	memory in cardiac tissue, Physical Review Letters 89~(13) (2002) 138101.
	
	\bibitem{cherry2004suppression}
	E.~M. Cherry, F.~H. Fenton, Suppression of alternans and conduction blocks
	despite steep {APD} restitution: {E}lectrotonic, memory, and conduction
	velocity restitution effects, American Journal of Physiology-Heart and
	Circulatory Physiology 286~(6) (2004) H2332--H2341.
	
	\bibitem{diaz2004sarcoplasmic}
	M.~E. D{\'\i}az, S.~C. O’neill, D.~A. Eisner, Sarcoplasmic reticulum calcium
	content fluctuation is the key to cardiac alternans, Circulation Research
	94~(5) (2004) 650--656.
	
	\bibitem{restrepo2009spatiotemporal}
	J.~G. Restrepo, A.~Karma, Spatiotemporal intracellular calcium dynamics during
	cardiac alternans, Chaos: An Interdisciplinary Journal of Nonlinear Science
	19~(3) (2009).
	
	\bibitem{pathak2018model}
	J.~Pathak, B.~Hunt, M.~Girvan, Z.~Lu, E.~Ott, Model-free prediction of large
	spatiotemporally chaotic systems from data: {A} {R}eservoir {C}omputing
	approach, Physical Review Letters 120~(2) (2018) 024102.
	
	\bibitem{ramadevi2022chaotic}
	B.~Ramadevi, K.~Bingi, Chaotic time series forecasting approaches using
	{M}achine {L}earning techniques: A review, Symmetry 14~(5) (2022) 955.
	
	\bibitem{may1976simple}
	R.~M. May, Simple mathematical models with very complicated dynamics, Nature
	261~(5560) (1976) 459--467.
	
	\bibitem{beeler1977reconstruction}
	G.~W. Beeler, H.~Reuter, Reconstruction of the action potential of ventricular
	myocardial fibres, The Journal of Physiology 268~(1) (1977) 177--210.
	
	\bibitem{paszke2019pytorch}
	A.~Paszke, S.~Gross, F.~Massa, A.~Lerer, J.~Bradbury, G.~Chanan, T.~Killeen,
	Z.~Lin, N.~Gimelshein, L.~Antiga, A.~Desmaison, A.~Kopf, E.~Yang, Z.~DeVito,
	M.~Raison, A.~Tejani, S.~Chilamkurthy, B.~Steiner, L.~Fang, J.~Bai,
	S.~Chintala, Py{T}orch: {A}n imperative style, high-performance {D}eep
	{L}earning library, Advances in Neural Information Processing Systems 32
	(2019).
	
	\bibitem{barrio2023deepdata}
	R.~Barrio, {\'A}.~Lozano, A.~Mayora-Cebollero, C.~Mayora-Cebollero, A.~Miguel,
	A.~Ortega, S.~Serrano, R.~Vigara,
	\href{https://data.mendeley.com/datasets/k4x675k5dm/3}{Dataset of article:
		"{D}eep {L}earning for chaos detection"} (2023).
	\newblock \href {https://doi.org/10.17632/k4x675k5dm.3}
	{\path{doi:10.17632/k4x675k5dm.3}}.
	\newline\url{https://data.mendeley.com/datasets/k4x675k5dm/3}
	
	\bibitem{hochreiter1997long}
	S.~Hochreiter, J.~Schmidhuber, Long {S}hort-{T}erm {M}emory, Neural computation
	9~(8) (1997) 1735--1780.
	
	\bibitem{redmon2016you}
	J.~Redmon, S.~Divvala, R.~Girshick, A.~Farhadi, You {O}nly {L}ook {O}nce:
	{U}nified, real-time object detection, in: Proceedings of the IEEE Conference
	on Computer Vision and Pattern Recognition, 2016, pp. 779--788.
	
	\bibitem{liu2019arrhythmias}
	F.~Liu, X.~Zhou, J.~Cao, Z.~Wang, H.~Wang, Y.~Zhang, Arrhythmias classification
	by integrating stacked bidirectional {LSTM} and two-dimensional {CNN}, in:
	Advances in Knowledge Discovery and Data Mining: 23rd Pacific-Asia
	Conference, PAKDD 2019, Macau, China, April 14-17, 2019, Proceedings, Part II
	23, Springer, 2019, pp. 136--149.
	
	\bibitem{engelken2024gradient}
	R.~Engelken, Gradient flossing: Improving gradient descent through dynamic
	control of jacobians, in: A.~Oh, T.~Naumann, A.~Globerson, K.~Saenko,
	M.~Hardt, S.~Levine (Eds.), Advances in Neural Information Processing
	Systems, Vol.~36, Curran Associates, Inc., 2023, pp. 10412--10439.
	
	\bibitem{cho2014learning}
	K.~Cho, B.~Van~Merri{\"e}nboer, C.~Gulcehre, D.~Bahdanau, F.~Bougares,
	H.~Schwenk, Y.~Bengio, Learning phrase representations using {RNN}
	encoder-decoder for statistical machine translation, arXiv preprint
	arXiv:1406.1078 (2014).
	
	\bibitem{kingma2014adam}
	D.~P. Kingma, J.~Ba, Adam: {A} method for stochastic optimization, arXiv
	preprint arXiv:1412.6980 (2014).
	
	\bibitem{Feigenbaum78}
	M.~J. Feigenbaum, Quantitative universality for a class of nonlinear
	transformations, Journal of Statistical Physics 19~(1) (1978) 25--52.
	
	\bibitem{vallat2025antropy}
	R.~Vallat, Antropy: Entropy and complexity of time-series in python (v0.1.9),
	\url{https://github.com/raphaelvallat/antropy} (2025).
	
	\bibitem{matthews2025chaos}
	P.~Matthews, 0\,-1 test for chaos,
	\url{https://www.mathworks.com/matlabcentral/fileexchange/25050-0-1-test-for-chaos},
	{MATLAB} Central File Exchange (2025).
	
	\bibitem{lewis1990chaotic}
	T.~J. Lewis, M.~R. Guevara, Chaotic dynamics in an ionic model of the
	propagated cardiac action potential, Journal of Theoretical Biology 146~(3)
	(1990) 407--432.
	
	\bibitem{cherry2008visualization}
	E.~M. Cherry, F.~H. Fenton, Visualization of spiral and scroll waves in
	simulated and experimental cardiac tissue, New Journal of Physics 10~(12)
	(2008) 125016.
	
	\bibitem{cherry2007tale}
	E.~M. Cherry, F.~H. Fenton, A tale of two dogs: {A}nalyzing two models of
	canine ventricular electrophysiology, American Journal of Physiology-Heart
	and Circulatory Physiology 292~(1) (2007) H43--H55.}}
	
\end{thebibliography}

\end{document}